\documentclass[journal,draftcls,onecolumn]{IEEEtran}

\usepackage{cite}
\usepackage[dvips]{graphicx}
\graphicspath{{./figures/}}
\DeclareGraphicsExtensions{.eps}
\usepackage[cmex10]{amsmath}
\usepackage{amssymb}
\usepackage{psfrag}
\usepackage{xspace}

\newcommand{\ie} {i.\,e.\xspace}
\newcommand{\eg} {e.\,g.\xspace}
\newcommand{\cf} {cf.\xspace}
\newcommand{\iid} {i.\,i.\,d.\xspace}

\DeclareMathOperator{\dB}{dB}
\DeclareMathOperator{\Besseli}{I}	
\newcommand{\mcal}{\mathcal}		
\newcommand{\vphi}{\varphi}
\newcommand{\iPhi}{{\it \Phi}}
\newcommand{\jm}{\jmath\,}
\newcommand{\Real}[1]{\Re\{#1\}}
\newcommand{\Imag}[1]{\Im\{#1\}}
\newcommand{\expectation}[2]{{\mcal E}_{#1}\left\{ #2 \right\}}
\newcommand{\dd}[1]{{\rm d}{#1}}	
\newcommand{\erfc}{\mbox{erfc}}
\newcommand{\sinc}[1]{{\rm sinc}\left(#1\right)}
\newcommand{\rect}[1]{{\rm rect}\left(#1\right)}
\newcommand{\sabs}[1]{{#1}_\shortparallel}		
\newcommand{\sang}[1]{{#1}_\sphericalangle}		
\newcommand{\degree}{\,^{\circ}}
\newcommand{\normalc}[2]{\mathcal{N}_\mathbb{C}\left( #1, #2 \right)}
\newcommand{\normalr}[2]{\mathcal{N}_\mathbb{R}\left( #1, #2 \right)}
\newcommand{\Fourier}[1]{\mathcal{F}\left({#1}\right)}		

\begin{document}
%
\title{Calculation of Mutual Information for Partially Coherent Gaussian Channels with Applications to Fiber Optics}
%
%
%

\author{Bernhard~Goebel,~\IEEEmembership{Member,~IEEE,}
        Ren\'e-Jean~Essiambre,~\IEEEmembership{Senior Member,~IEEE,}
        Gerhard~Kramer,~\IEEEmembership{Fellow,~IEEE,}
        Peter~J.~Winzer,~\IEEEmembership{Fellow,~IEEE,}
        and~Norbert~Hanik,~\IEEEmembership{Member,~IEEE}
\thanks{B.~Goebel, G.~Kramer and N.~Hanik are with the Institute for Communications Engineering, Technische Universit\"at M\"unchen, 80290 Munich, Germany (e-mail: bernhard.goebel@tum.de).}
\thanks{R.-J.~Essiambre and P.~J.~Winzer are with Alcatel-Lucent, Bell Labs, Holmdel, NJ 07733, USA.}
}


\maketitle

\begin{abstract}
The mutual information between a complex-valued channel input and its complex-valued output is decomposed into four parts based on polar coordinates: an amplitude term, a phase term, and two mixed terms.
Numerical results for the additive white Gaussian noise (AWGN) channel with various inputs show that, at high signal-to-noise ratio (SNR), the amplitude and phase terms dominate the mixed terms.
For the AWGN channel with a Gaussian input, analytical expressions are derived for high SNR.
The decomposition method is applied to partially coherent channels and a property of such channels called ``spectral loss'' is developed.
Spectral loss occurs in nonlinear fiber-optic channels and it may be one effect that needs to be taken into account to explain the behavior of the capacity of nonlinear fiber-optic channels presented in recent studies.
\end{abstract}


\section{Introduction}
\IEEEPARstart{T}{he} information encoded in complex-valued signals has two degrees of freedom which are commonly taken to be the signal's two quadratures -- its real and imaginary parts.
Alternatively, the signal can be decomposed into its polar coordinates -- amplitude and phase.
Historically, the first digital modulation constellations with two degrees of freedom were a combination of one-dimensional amplitude modulation (AM) and phase modulation (PM)~\cite{Cahn1960}.
Quadrature amplitude modulation (QAM), \ie, amplitude modulation of two orthogonal carriers, was not described until 1962, with the most significant progress in understanding made in the 1970s~\cite[Sec.~1.2]{Hanzo_QAM}.

The decomposition of complex-valued signals into their real and imaginary parts is the method of choice when the sub-channels transporting them have identical form and noise statistics.
In particular, this is the case for the AWGN channel, \eg, with circularly symmetric Gaussian or square QAM input.
In contrast, the ``old-fashioned'' AM-PM view can be useful when physical effects act differently on the different sub-channels.
Examples are systems that clip the amplitude (\eg, nonlinear amplifiers) or systems that introduce phase noise (\eg, phase-locked loops or certain nonlinear optical fiber effects).
However, even for channels that introduce equal impairments to the signal's quadratures (such as the complex-valued AWGN channel), the AM-PM view may be preferable if this facilitates the input description, for instance for ASK-PSK modulation schemes.

Decomposing signals using polar coordinates motivates decomposing the mutual information between the channel input and output using polar coordinates.
We choose a decomposition that results in four terms:
two partial channels, each with one degree of freedom (an amplitude and a phase channel), and two mixed terms that govern the exchange of mutual information across the sub-channels.
We explain and discuss this method in Section~\ref{sec:decomposition}.
We illustrate our results by applying the decomposition to the complex-valued AWGN channel.
In Section~\ref{sec:AWGN}, we derive analytical expressions for the AWGN channel with average power constraint (Gaussian input) and with constant power constraint (phase-modulated input).
In addition, we present decomposition results for discrete ASK/PSK and QAM constellations.

The second part of the paper deals with partially coherent channels, which are essentially AWGN channels with additional phase noise.
Such channels motivate the development of the polar decomposition method described in this paper.
The earliest information-theoretic results on channels with reduced degrees of freedom, \eg, transmitters or receivers that are limited to amplitude modulation (AM) or phase modulation (PM), date back to 1953~\cite{Blachman1953}.
Some time later, partially coherent channels became an important research topic in the context of phase jitter induced by phase demodulation~\cite{Viterbi1963}.
Good modulation schemes for such channels were presented in~\cite{Foschini1973}.
To this date, little is known about the capacity-achieving input for partially coherent channels~\cite{Katz2004}.
We discuss partially coherent channels in Section~\ref{sec:partially_coherent} and derive an effect we call ``spectral loss''.
At the end of that section, we use the capacity of fiber-optic communication channels as one application of our results.
Finally, in Appendix~\ref{app:directional_statistics} we review results from directional statistics that are useful for understanding phase noise and other circular random processes.

\section{A Polar Decomposition of Mutual Information}\label{sec:decomposition}

Consider a channel with complex-valued input 
\begin{equation}\label{eq:complex_input}
X = \sabs{X} \cdot e^{\jm \sang{X}}, \quad \sabs{X} \in [0, \infty), \sang{X} \in [-\pi, \pi )
\end{equation}
and output
\begin{equation}\label{eq:complex_output}
Y = \sabs{Y} \cdot e^{\jm \sang{Y}}, \quad \sabs{Y} \in [0, \infty), \sang{Y} \in [-\pi, \pi ),
\end{equation}
where the notation $\sabs{X}, \sabs{Y}$ (amplitudes) and $\sang{X}, \sang{Y}$ (phase angles) reminds us of what parts of the signal these variables refer to.
(We use lower-case fonts $\sabs{x}$ to denote a realization and calligraphic fonts $\sabs{\mcal{X}}$ to denote the support of the random variable $\sabs{X}$.)

The mutual information $I(X;Y)$ between this channel's input and output can be expanded by repeatedly applying the chain rule of mutual information~\cite[p.~22]{CoverThomas} as
\begin{align}\label{eq:MI_polar_decomposition}
I(X;Y) &= I(\sabs{X}, \sang{X}; \sabs{Y}, \sang{Y}) \nonumber \\
  &= I(\sabs{X}; \sabs{Y}, \sang{Y}) + I(\sang{X}; \sabs{Y},\sang{Y} | \sabs{X}) \nonumber \\
  &= \underbrace{I(\sabs{X}; \sabs{Y})}_{\mathrm{Amplitude ~ term}} +
     \underbrace{I(\sang{X}; \sang{Y} | \sabs{X})}_{\mathrm{Phase ~ term}} +
     \underbrace{I(\sabs{X}; \sang{Y} | \sabs{Y})}_{\mathrm{Mixed ~ term ~ I}} +
     \underbrace{I(\sang{X}; \sabs{Y} | \sabs{X},\sang{Y})}_{\mathrm{Mixed ~ term ~ II}}.
\end{align}

The expansion (\ref{eq:MI_polar_decomposition}) can be interpreted as decomposing the complex-valued channel with two degrees of freedom (amplitude and phase) into two sub-channels with one degree of freedom each.

The first sub-channel, represented by the amplitude term of the mutual information
\begin{equation}\label{eq:MI_decomposition_amplitude_term}
I(\sabs{X};\sabs{Y}) = \int_{\sabs{\mcal{X}}} p(\sabs{x}) \int_{\sabs{\mcal{Y}}} p(\sabs{y}|\sabs{x}) \log\frac{p(\sabs{y}|\sabs{x})}{p(\sabs{y})} \dd{\sabs{y}}\dd{\sabs{x}}
\end{equation}
conveys only the amplitude of the signal and is unaffected by impairments such as phase noise.

The second sub-channel is characterized by the phase term of the mutual information
\begin{align}\label{eq:MI_decomposition_phase_term}
I(\sang{X}; \sang{Y} | \sabs{X}) &=
     \int_{\sabs{\mcal{X}}} p(\sabs{x}) I(\sang{X}; \sang{Y} | \sabs{x}) \dd{\sabs{x}} \nonumber \\
  &= \int_{\sabs{\mcal{X}}} p(\sabs{x})
  	\iint_{\sang{\mcal{X}},\sang{\mcal{Y}}} p(\sang{x},\sang{y}|\sabs{x})
  	\log \frac{p(\sang{x},\sang{y}|\sabs{x})}{p(\sang{x}|\sabs{x}), p(\sang{y}|\sabs{x})}
  	\dd{\sang{x}} \dd{\sang{y}} \dd{\sabs{x}} \nonumber \\
  &= \int_{\sabs{\mcal{X}}} p(\sabs{x})
		\underbrace{
			\int_{\sang{\mcal{X}}} p(\sang{x}|\sabs{x})
			\int_{\sang{\mcal{Y}}} p(\sang{y}|\sang{x},\sabs{x})
			\log \frac{p(\sang{y}|\sang{x},\sabs{x})}{p(\sang{y}|\sabs{x})}
			\dd{\sang{y}} \dd{\sang{x}}
		}_{I(\sang{X};\sang{Y}|\sabs{x})}
  	\dd{\sabs{x}} \nonumber \\
  &= \expectation{\sabs{X}}{I(\sang{X};\sang{Y}|\sabs{X}=\sabs{x})},
\end{align}
where $\expectation{X}{f(X\!=\!x)}$ denotes the expectation of $f(X)$ with respect to the random variable $X$ that takes on the values $x$.
Eq.~(\ref{eq:MI_decomposition_phase_term}) can be paraphrased in words as the information that can be obtained about the \emph{input phase} by observing the \emph{output phase}, given that the \emph{input amplitude} is already known.
This term is significantly affected by phase noise, but agnostic to amplitude distortions such as clipping as long as the input amplitude is known.

After separating the complex-valued channel into an amplitude and a phase part, the two mixed terms (I and II) in (\ref{eq:MI_polar_decomposition}) yield the ``cross information'' between these two sub-channels.
Mixed term~I represents the amount of information about the \emph{input amplitude} that can be drawn from the \emph{output phase} in addition to what has already been learnt about the \emph{input amplitude} by observing the \emph{output amplitude}.
Finally, mixed term~II yields the information about the \emph{input phase} that can be obtained from observation of the \emph{output amplitude} given the \emph{input amplitude} and the \emph{output phase}.

The polar decomposition of mutual information can be helpful in understanding the characteristics of the channel input, \eg, concerning symbol constellations, and transmission impairments.
Moreover, the decomposition significantly simplifies the computation of the mutual information in cases where the mixed terms are zero or negligibly small.
The computation of $I(X;Y)$ then reduces to evaluating the conditional probability densities in (\ref{eq:MI_decomposition_amplitude_term}) and (\ref{eq:MI_decomposition_phase_term}), which are often known.
Even if the mixed terms do not vanish, the two main terms yield a lower bound on the mutual information (and can hence be used to get a lower bound on capacity).


\section{Decomposition of the AWGN Channel}\label{sec:AWGN}
We next apply the decomposition (\ref{eq:MI_polar_decomposition}) to the complex-valued AWGN channel
\begin{equation}
Y = X + N, \quad N \sim \normalc{0}{2\sigma_{\rm n}^2},
\label{eq:AWGN_definition}
\end{equation}
with the power constraint $\expectation{}{|X|^2} \leq P_{\rm s}$.
This channel's signal-to-noise ratio (SNR), stated here for later reference, is
\begin{equation}\label{eq:SNR}
\frac{P_{\rm s}}{2\sigma_{\rm n}^2}
\end{equation}
or (in dB)
\begin{equation}\label{eq:SNRdB}
10 \cdot \log_{10}\left( \frac{P_{\rm s}}{2\sigma_{\rm n}^2} \right).
\end{equation}

\subsection{Gaussian Input}\label{subsec:AWGN_Gaussian}
\subsubsection{Amplitude term}
The first term in the decomposition is $I(\sabs{X};\sabs{Y}) = h(\sabs{Y})-h(\sabs{Y}|\sabs{X})$.
The capacity of the AWGN channel (\ref{eq:AWGN_definition}) with average power constraint is maximized by $X \sim \normalc{0}{P_{\rm s}}$~\cite[p.~242]{CoverThomas}.
Since $N \sim \normalc{0}{2\sigma_{\rm n}^2}$, the channel output is Gaussian distributed, $Y \sim \normalc{0}{P_{\rm s}+2\sigma_{\rm n}^2}$.
Then, $\sabs{Y} = \sqrt{\Real{Y}^2+\Imag{Y}^2}$ follows a Rayleigh distribution with parameter $(P_{\rm s}+2\sigma_{\rm n}^2)/2$~\cite[p.~45]{Proakis}:
\begin{equation}\label{eq:Rayleigh_distribution}
p(\sabs{y}) = \frac{2 \sabs{y}}{P_{\rm s}+2\sigma_{\rm n}^2} \cdot \exp \left( -\frac{\sabs{y}^2}{P_{\rm s}+2\sigma_{\rm n}^2} \right).
\end{equation}
The differential entropy of the output amplitude in bits is~\cite[p.~487]{CoverThomas}
\begin{equation}\label{eq:diff_entropy_output_amplitude}
h(\sabs{Y}) = \frac{1}{2} \log_2 (P_{\rm s}+2\sigma_{\rm n}^2) + \frac{1}{\ln 2} + \frac{\gamma}{2\ln 2} - 1,
\end{equation}
where $\gamma \approx 0.577$ is the \emph{Euler constant}.

Calculating $h(\sabs{Y}|\sabs{X})$ requires knowledge of $p(\sabs{y}|\sabs{x})$, which is a Ricean distribution~\cite[p.~47]{Proakis}:
\begin{equation}\label{eq:Ricean_distribution}
p(\sabs{y}|\sabs{x}) = \frac{\sabs{y}}{\sigma_{\rm n}^2} \cdot
			\exp \left( -\frac{\sabs{x}^2+\sabs{y}^2}{2\sigma_{\rm n}^2} \right) \cdot
			\Besseli_0 \left( \frac{\sabs{x}\sabs{y}}{\sigma_{\rm n}^2} \right),
\end{equation}
where $\Besseli_0(.)$ is the modified Bessel function of the first kind with order zero.
It can be seen that for $\sabs{x} \!=\! 0$, the Ricean distribution turns into a Rayleigh distribution; for $P_{\rm s}=0$, (\ref{eq:Rayleigh_distribution}) and (\ref{eq:Ricean_distribution}) are equal.
Using the general form (\ref{eq:Ricean_distribution}) of the conditional PDF, the integration required to calculate $h(\sabs{Y}|\sabs{X})$ is intractable.
A significant simplification is obtained when the channel's SNR (\ref{eq:SNR}) is large.
In this limit of large arguments of the Bessel function ($\sabs{x}\sabs{y}/\sigma_{\rm n}^2 \gg 1$), we can use $\Besseli_0(z) \to e^z/\sqrt{2 \pi z}$~\cite[p.~377]{Abramowitz}.
The Ricean PDF (\ref{eq:Ricean_distribution}) then turns into the Gaussian PDF
\begin{equation}\label{eq:Ricean_limit_Gaussian}
p(\sabs{y}|\sabs{x}) \approx \frac{1}{\sigma_{\rm n} \sqrt{2\pi}} \cdot
			\exp \left( -\frac{(\sabs{y}-\sabs{x})^2}{2\sigma_{\rm n}^2} \right).
\end{equation}
In deriving (\ref{eq:Ricean_limit_Gaussian}), we dropped a factor $\sqrt{\sabs{y}/\sabs{x}}$ which decays to $1$ asymptotically with increasing SNR.
With (\ref{eq:Ricean_limit_Gaussian}), the conditional differential entropy can be calculated as
\begin{equation}\label{eq:diff_entropy_output_amplitude_given_input_amplitude}
h(\sabs{Y}|\sabs{X}) \approx \frac{1}{2} \log_2 (2 \pi e \sigma_{\rm n}^2).
\end{equation}
Finally, using (\ref{eq:diff_entropy_output_amplitude_given_input_amplitude}) and (\ref{eq:diff_entropy_output_amplitude}), an asymptotic approximation for the amplitude term is
\begin{align}\label{eq:AWGN_amplitude_term_asymptotic}
I(\sabs{X};\sabs{Y}) &=  h(\sabs{Y})-h(\sabs{Y}|\sabs{X}) \nonumber \\
	&\approx \frac{1}{2} \log_2 \left( 1+ \frac{P_{\rm s}}{2\sigma_{\rm n}^2} \right) -
			\frac{1}{2} \log_2 \pi + \frac{1+\gamma}{2\ln 2} - 1 \nonumber \\
	&\approx \frac{1}{2} \log_2 \frac{P_{\rm s}}{2\sigma_{\rm n}^2} \underbrace{-
			\frac{1}{2} \log_2 \pi + \frac{1+\gamma}{2\ln 2} - 1}_{\approx -0.69}, \qquad P_{\rm s} \gg 2\sigma_{\rm n}^2.
\end{align}

\subsubsection{Phase term}
The phase term $I(\sang{X}; \sang{Y} | \sabs{X}) = h(\sang{Y} | \sabs{X}) - h(\sang{Y} | \sang{X},\sabs{X})$ is calculated similarly.
For any input amplitude $\sabs{x}$, the output phase is uniformly distributed in $[-\pi, \pi)$, so the first conditional entropy is easily found to be
\begin{align}\label{eq:diff_entropy_output_phase_given_input_amplitude}
h(\sang{Y} | \sabs{X}) &= - \int_{\sabs{\mcal{X}}} \int_{\sang{\mcal{Y}}}
		p(\sabs{x},\sang{y}) \log_2 p(\sang{y}|\sabs{x}) \dd{\sang{y}}\dd{\sabs{x}} \nonumber \\
	&= - \int_{\sabs{\mcal{X}}} p(\sabs{x}) \underbrace{\int_{-\pi}^{\pi} p(\sang{y} | \sabs{x})
		\log_2 p(\sang{y}|\sabs{x}) \dd{\sang{y}}}_{-\log_2(2\pi)} \dd{\sabs{x}} \nonumber \\
	&= \log_2(2\pi).
\end{align}
Similarly, we can write
\begin{equation}\label{eq:diff_entropy_output_phase_given_complexinput}
h(\sang{Y} | \sang{X},\sabs{X}) =
	- \iint_{\sabs{\mcal{X}},\sang{\mcal{X}}} p(\sabs{x},\sang{x}) \cdot
	\underbrace{\int_{-\pi}^{\pi} p(\sang{y} | \sabs{x},\sang{x})
	\log_2 p(\sang{y} | \sabs{x},\sang{x}) \dd{\sang{y}}}_{-h(\sang{Y}|\sabs{x},\sang{x})}
	\dd{\sabs{x}}\dd{\sang{x}}.
\end{equation}
The conditional entropy $h(\sang{Y}|\sabs{x},\sang{x})$ is not affected by the constant phase shift $\sang{x}$, so that we can assume $\sang{x} \!=\! 0$ without loss of generality and write the conditional phase PDF as~\cite{Middleton,Proakis,Aldis1993}
\begin{align}\label{eq:phase_PDF_AWGN}
p(\sang{y}|\sabs{x},\sang{x}=0) = & \frac{1}{2\pi} \exp\left( -\frac{\sabs{x}^2}{2\sigma_{\rm n}^2} \right) \nonumber \\
	&+ \frac{\sabs{x}\cos\sang{y}}{2\sqrt{\pi 2\sigma_{\rm n}^2}} \cdot
			\exp\left( -\frac{\sabs{x}^2(\sin\sang{y})^2}{2\sigma_{\rm n}^2} \right) \cdot
			\erfc\left( - \frac{\sabs{x}\cos\sang{y}}{\sqrt{2\sigma_{\rm n}^2}} \right).
\end{align}
The PDF (\ref{eq:phase_PDF_AWGN}) is periodic with a period of $2\pi$; integrating it over any contiguous $2\pi$ interval yields one.
Such \emph{circular} PDFs are reviewed in Appendix~\ref{app:directional_statistics}.

If the channel SNR is low and we have $\sabs{x}^2 \ll 2\sigma_{\rm n}^2$, the phase becomes uniformly distributed.
On the other hand, when $\sabs{x}^2 \gg 2\sigma_{\rm n}^2$, (\ref{eq:phase_PDF_AWGN}) can be approximated by the Gaussian PDF~\cite[p.~273]{Proakis}
\begin{equation}\label{eq:phase_PDF_AWGN_limit_Gaussian}
p(\sang{y}|\sabs{x},\sang{x}=0) \approx \frac{\sabs{x}}{\sqrt{2\pi}\sigma_{\rm n}} \cdot
	\exp\left(- \frac{\sang{y}^2}{2\sigma_{\rm n}^2/\sabs{x}^2} \right).
\end{equation}
With this approximation, the inner entropy integral in (\ref{eq:diff_entropy_output_phase_given_complexinput}) can be approximated as
\begin{equation}\label{eq:diff_entropy_output_phase_given_complexinput_realization}
h(\sang{Y}|\sabs{x},\sang{x}) \approx \frac{1}{2} \cdot \log_2 \left( 2 \pi e \cdot \frac{\sigma_{\rm n}^2}{\sabs{x}^2} \right), \qquad \sabs{x}^2 \gg 2\sigma_{\rm n}^2,
\end{equation}
and the entropy (\ref{eq:diff_entropy_output_phase_given_complexinput}) becomes
\begin{align}\label{eq:diff_entropy_output_phase_given_complexinput_solution}
h(\sang{Y} | \sang{X},\sabs{X}) &=
		\int_{\sang{\mcal{X}}} p(\sang{x}) \cdot \int_{\sabs{\mcal{X}}} p(\sabs{x}|\sang{x}) \cdot
		h(\sang{Y}|\sabs{x},\sang{x}) \dd{\sabs{x}} \dd{\sang{x}} \nonumber \\
	&= \int_{-\pi}^{\pi} p(\sang{x}) \dd{\sang{x}} \cdot \int_{0}^{\infty} p(\sabs{x}) \cdot
		h(\sang{Y}|\sabs{x},\sang{x}) \dd{\sabs{x}} \nonumber \\
	&\approx \int_{0}^{\infty}
		\frac{\sabs{x}}{P_{\rm s}/2} \cdot \exp \left( -\frac{\sabs{x}^2}{P_{\rm s}} \right) \cdot
		\frac{1}{2} \log_2 \left( 2 \pi e \cdot \frac{\sigma_{\rm n}^2}{\sabs{x}^2} \right)
		\dd{\sabs{x}} \nonumber \\
	&= \frac{1}{2} \log_2 \frac{2\sigma_{\rm n}^2}{P_{\rm s}} +
		\frac{1 + \gamma}{2 \ln 2} +
		\frac{1}{2} \log_2 \pi, \qquad P_{\rm s} \gg 2\sigma_{\rm n}^2.
\end{align}
The separation of the integrals (second equality) is possible because $h(\sang{Y}|\sabs{x},\sang{x})$ is independent of $\sang{x}$.
In the same line, we used $p(\sabs{x}|\sang{x}) \!=\! p(\sabs{x})$, which is a Rayleigh distribution.

Finally, the decomposition phase term can be approximated from (\ref{eq:diff_entropy_output_phase_given_input_amplitude}) and (\ref{eq:diff_entropy_output_phase_given_complexinput_solution}):
\begin{align}\label{eq:AWGN_phase_term_asymptotic}
I(\sang{X}; \sang{Y} | \sabs{X}) &= h(\sang{Y} | \sabs{X}) - h(\sang{Y} | \sang{X},\sabs{X}) \nonumber \\
	&\approx \log_2(2\pi) - \frac{1}{2} \log_2 \frac{2\sigma_{\rm n}^2}{P_{\rm s}} -
		\frac{1 + \gamma}{2 \ln 2} - \frac{1}{2} \log_2 \pi \nonumber \\
	&= \frac{1}{2} \log_2 \frac{P_{\rm s}}{2\sigma_{\rm n}^2} + \frac{1}{2} \log_2 \pi -
		\frac{1+\gamma}{2\ln 2} + 1, \qquad P_{\rm s} \gg 2\sigma_{\rm n}^2.
\end{align}

\subsubsection{Mixed terms}

For the AWGN channel with Gaussian input, \emph{mixed term I} in the decomposition is always zero.
To prove this, observe that $p(\sang{y}) \!=\! p(\sang{y}|\sabs{x}) \!=\! p(\sang{y}|\sabs{x},\sabs{y}) \!=\!
1/(2\pi)$ within any $2\pi$ interval and zero outside.
Then, we obtain the conditional entropies
\begin{align}\label{eq:decomposition_MTI_entropy1}
h(\sang{Y} | \sabs{Y}) &= - \int_{0}^{\infty} p(\sabs{x})
	\underbrace{\int_{-\pi}^{\pi} p(\sang{y}|\sabs{x}) \log_2 p(\sang{y}|\sabs{x}) \dd{\sang{y}}}_{-\log_2(2\pi)}
	\dd{\sabs{x}}
	= \int_{0}^{\infty} p(\sabs{x}) \dd{\sabs{x}} \cdot \log_2(2\pi) \nonumber \\
 &= \log_2(2\pi)
\end{align}
and
\begin{align}\label{eq:decomposition_MTI_entropy2}
h(\sang{Y} | \sabs{Y},\sabs{X}) &= - \iint_{0}^{\infty} p(\sabs{x},\sabs{y})
	\underbrace{\int_{-\pi}^{\pi} p(\sang{y}|\sabs{x},\sabs{y}) \log_2 p(\sang{y}|\sabs{x},\sabs{y}) \dd{\sang{y}}}_{-\log_2(2\pi)} \dd{\sabs{x}} \dd{\sabs{y}} \nonumber \\
 &= \iint_{0}^{\infty} p(\sabs{x},\sabs{y}) \dd{\sabs{x}} \dd{\sabs{y}} \cdot \log_2(2\pi)
	= \log_2(2\pi),
\end{align}
and so
\begin{equation}\label{eq:decomposition_mixed_term_I}
I(\sabs{X}; \sang{Y} | \sabs{Y}) = h(\sang{Y} | \sabs{Y}) - h(\sang{Y} | \sabs{Y},\sabs{X}) = 0.
\end{equation}

\emph{Mixed term II}, $I(\sang{X}; \sabs{Y} | \sabs{X},\sang{Y})$, reaches its (numerically calculated) maximum value of approximately 0.08 bits/symbol at $10\log_{10}(P_{\rm s}/(2\sigma_{\rm n}^2)) \!=\! 1\dB$ and tends to zero for large SNRs.

\begin{figure}[!htb]
\centering
  \psfrag{C}[][]{Mutual information in bits per symbol}
  \psfrag{SNR}[][]{SNR in dB}
  \psfrag{Legend1xxxxxxxxx}{\small $I(X;Y)$}
  \psfrag{Legend2xxxxxxxxx}{\small $I(\sabs{X};\sabs{Y})$}
  \psfrag{Legend3xxxxxxxxx}{\small $I(\sang{X}; \sang{Y} | \sabs{X})$}
  \psfrag{Legend4xxxxxxxxx}{\small $I(\sang{X}; \sabs{Y} | \sabs{X},\sang{Y})$}
  \includegraphics[width=0.7\textwidth]{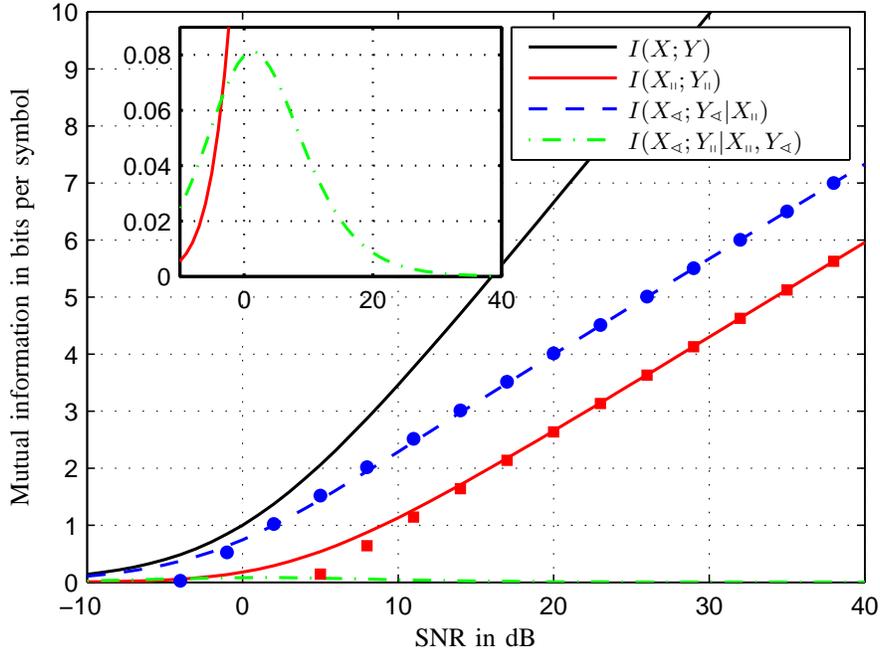}
\caption{Mutual information decomposition terms as a function of SNR in dB (\ref{eq:SNRdB}) for the AWGN channel with Gaussian input. Lines show numerical results, markers correspond to analytical approximations (\ref{eq:AWGN_amplitude_term_asymptotic}) and (\ref{eq:AWGN_phase_term_asymptotic}). The inset shows the magnified curve of mixed term II.}\label{fig:decomposition_Gaussian}
\end{figure}

The results of the decomposition for the AWGN channel with Gaussian input are shown in Fig.~\ref{fig:decomposition_Gaussian}.
The depicted curves were obtained from numerical integration of the mutual information integrals; markers indicate the analytical approximations (\ref{eq:AWGN_amplitude_term_asymptotic}) and (\ref{eq:AWGN_phase_term_asymptotic}).
Observe that the amplitude and phase terms are the main contributors to the channel capacity, whereas mixed term~II (shown in the inset) is negligibly small.
It can be seen that the analytical approximations are accurate at SNRs of approximately 15~dB and higher.
At high SNRs, both mixed terms are (exactly or near) zero and the amplitude and phase terms add up to the full capacity, as expected from (\ref{eq:AWGN_amplitude_term_asymptotic}) and (\ref{eq:AWGN_phase_term_asymptotic}).

It is noteworthy that the complex Gaussian input, which maximizes $I(X;Y)$, does not maximize the single decomposition terms independently.
The amplitude term $I(\sabs{X};\sabs{Y})$, for instance, is maximized by a ``half-Gaussian'' rather than a Rayleigh distribution at large SNRs, see Section~\ref{subsec:noncoherent}.

\subsection{Phase-modulated Input}\label{subsec:AWGN_PSK}
The terms \emph{constant-intensity}, \emph{constant-envelope} or \emph{ring} modulation are used in the literature to characterize the input of a system which encodes information only in the phase of the transmitted signal.
Results on the capacity of constant-intensity channels in the presence of AWGN have been reported over a period of 50 years, \eg,\cite{Blachman1953, Blahut1987, Geist1990, Aldis1993, HoKahn2002}.
The capacity of a channel constrained to constant intensity (``continuous PSK'') is an upper limit on the rates achievable with discrete PSK constellations.

An important detail in the definition of phase-modulated AWGN channels is whether the receiver has access to amplitude and phase of the received signal or to the phase only.
Although it has been observed~\cite{Aldis1993} that both capacities are equal in the limit of large SNRs, evaluating the capacity difference at lower SNR values has remained an open problem.

Performing a polar decomposition (\ref{eq:MI_polar_decomposition}) of the phase-modulated AWGN channel is the key to shed light on this question.
As no information is encoded in $\sabs{X} = \sqrt{P_{\rm s}} = {\rm const.}$, the amplitude term and the mixed term~I of the decomposition equal zero.

As expected for a phase-modulated system, the phase term conveys the greatest share of the transmitted information.
In the absence of amplitude modulation, this term can be written as $I(\sang{X};\sang{Y}|\sabs{X}\!=\!\sqrt{P_{\rm s}}) = h(\sang{Y}) - h(\sang{Y}|\sang{X})$.
The capacity-achieving input distribution is uniform in $[-\pi, \pi)$~\cite{Blahut1987}.
Hence, $\sang{Y}$ is uniformly distributed, too, and $h(\sang{Y}) = \log_2(2\pi)$.
To calculate $h(\sang{Y}|\sang{X})$, the entropy integral has to be solved for the conditional phase PDF (\ref{eq:phase_PDF_AWGN}).
An asymptotic approximation can be found for large SNRs, where (\ref{eq:phase_PDF_AWGN}) can be replaced by its Gaussian approximation (\ref{eq:phase_PDF_AWGN_limit_Gaussian}).
The conditional differential entropy $h(\sang{Y}|\sang{X})$ then approaches (\ref{eq:diff_entropy_output_phase_given_complexinput_realization}), and the decomposition phase term becomes
\begin{align}
I(\sang{X};\sang{Y}|\sabs{X}=\sqrt{P_{\rm s}})
 &= h(\sang{Y}) - h(\sang{Y}|\sang{X}) \nonumber \\
 &\approx \log_2(2\pi) - \frac{1}{2} \cdot \log_2 \left( 2 \pi e \cdot \frac{\sigma_{\rm n}^2}{P_{\rm s}} \right) \nonumber \\
 &= \frac{1}{2} \cdot \log_2 \left( \frac{4\pi}{e} \cdot \frac{P_{\rm s}}{2\sigma_{\rm n}^2} \right) \nonumber \\
 &\approx \frac{1}{2} \cdot \log_2 \frac{P_{\rm s}}{2\sigma_{\rm n}^2} + 1.1 \textrm{~bits}, \quad P_{\rm s} \gg 2\sigma_{\rm n}^2.
\label{eq:phase_only_decomposition_phase_term}
\end{align}
Hence, the capacity of the phase-modulated AWGN channel is approximately 1.1~bits/symbol larger than half that of the AWGN channel with Gaussian input for large SNRs.

\begin{figure}[!htb]
\centering
  \psfrag{C}[][]{Mutual information in bits per symbol}
  \psfrag{SNR}[][]{SNR in dB}
  \psfrag{Legend1xxxxxxxxx}{\small $I(\sang{X}; \sang{Y} | \sabs{X})$}
  \psfrag{Legend2xxxxxxxxx}{\small $I(\sang{X}; \sabs{Y} | \sabs{X},\sang{Y})$}
  \includegraphics[width=0.7\textwidth]{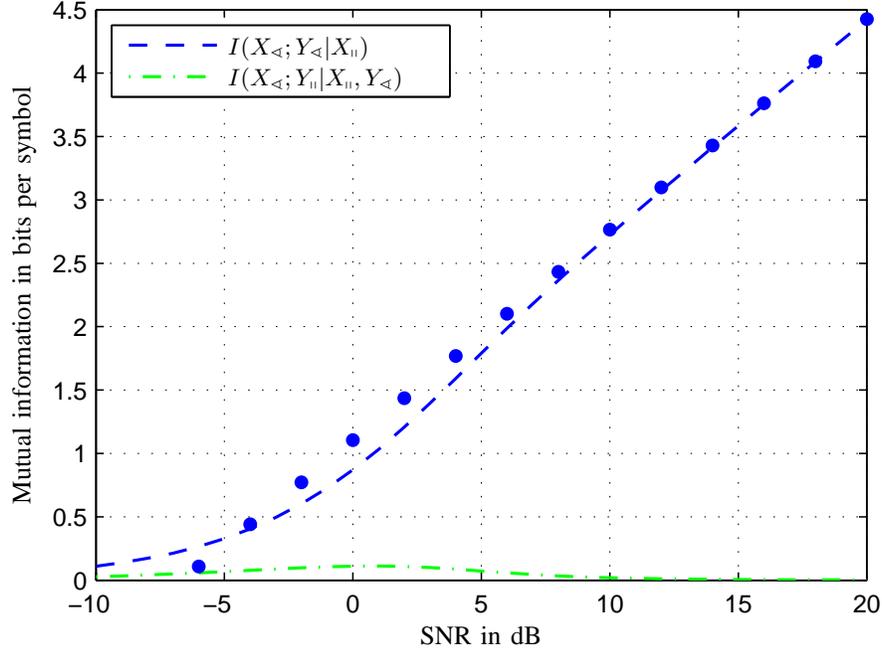}
\caption{Mutual information decomposition terms as a function of SNR in dB (\ref{eq:SNRdB}) for the AWGN channel with constant-intensity (continuous ring) input. Lines show numerical results, markers correspond to analytical approximation (\ref{eq:phase_only_decomposition_phase_term}).}\label{fig:decomposition_continuous_ring_1}
\end{figure}

Finally, mixed term~II, $I(\sang{X};\sabs{Y}|\sabs{X}\!=\!\sqrt{P_{\rm s}},\sang{Y})$, represents the (small) amount of information that can be gained by receiving the signal amplitude and phase rather than the phase only.
Fig.~\ref{fig:decomposition_continuous_ring_1} shows the decomposition terms as a function of SNR; the phase term markers indicate the asymptotic approximation~(\ref{eq:phase_only_decomposition_phase_term}), which is accurate at SNRs greater than 15~dB.

\subsection{Discrete Input Constellations}\label{subsec:AWGN_discrete}

In practical communication systems, the input consists of points from a discrete alphabet rather than of continuous values.
Performing the polar decomposition for these discrete inputs is useful in two ways:
\begin{itemize}
	\item
The decomposition can help to adapt constellations to certain channel characteristics.
For example, it may be beneficial for channels impaired by strong phase noise to re-arrange the points of a constellation in a way that the amplitude term is increased at the expense of the phase term.
While the overall capacity may be hardly affected in the absence of phase noise, an increased capacity is obtained in the presence of phase noise.
An example for this situation can be found in \cite{Belzer2002}, where 8-PSK is compared to 8-OOK-PSK (7-PSK plus a point at the origin) and 8-star-QAM in the presence of fading and phase noise.
The decomposition could help to accelerate this search for good constellations and possibly to make it more systematic.
	\item
When determining the mutual information numerically, the computational complexity can be significantly reduced by calculating the amplitude and phase terms rather than the full mutual information.
This approach requires both mixed terms to be negligibly small.
\end{itemize}
In the following, decomposition results are given for some exemplary modulation schemes.

\subsubsection{Modulation using one degree of freedom}

\begin{figure}[!htb]
\centering
  \psfrag{C}[][]{\footnotesize Mutual information in bits per symbol}
  \psfrag{SNR}[][]{\footnotesize SNR in dB}
  \psfrag{Legend1xxxxxxxxx}{\scriptsize $I(X;Y)$}
  \psfrag{Legend2xxxxxxxxx}{\scriptsize $I(\sabs{X};\sabs{Y})$}
  \psfrag{Legend3xxxxxxxxx}{\scriptsize $I(\sang{X}; \sang{Y} | \sabs{X})$}
  \psfrag{Legend4xxxxxxxxx}{\scriptsize $I(\sabs{X}; \sang{Y} | \sabs{Y})$}
  \psfrag{Legend5xxxxxxxxx}{\scriptsize $I(\sang{X}; \sabs{Y} | \sabs{X},\sang{Y})$}
  \includegraphics[width=0.48\textwidth]{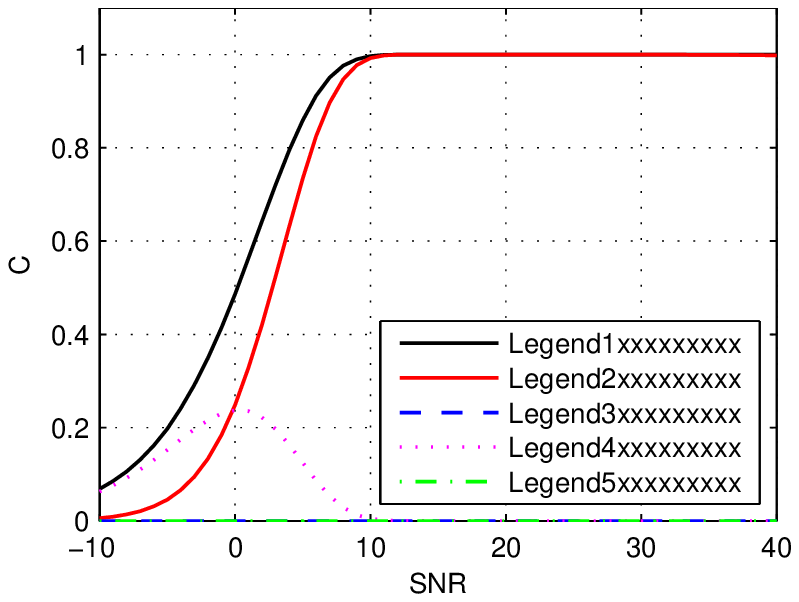}
  \includegraphics[width=0.48\textwidth]{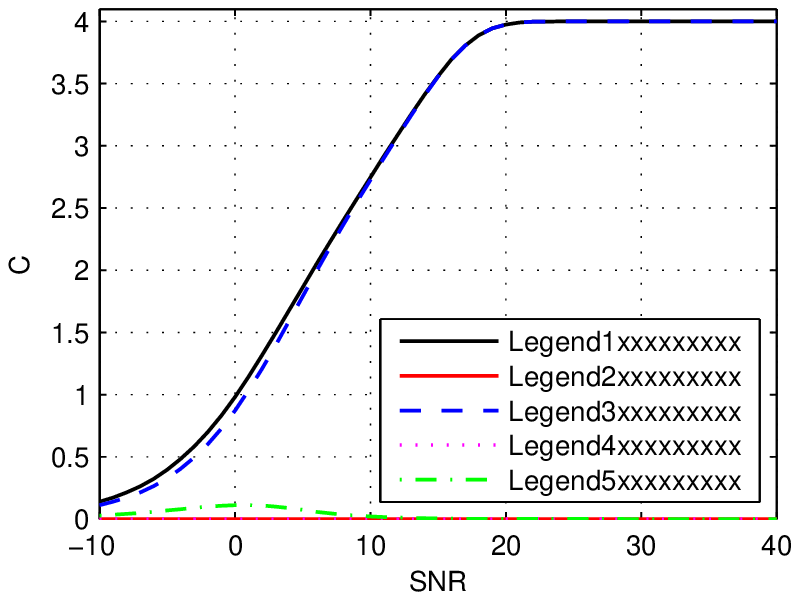}
\caption{Polar decomposition of mutual information for OOK (left) and 16-PSK (right).}\label{fig:decomposition_OOK_and_PSK}
\end{figure}

As examples of modulation schemes where either amplitude or phase are modulated, Fig.~\ref{fig:decomposition_OOK_and_PSK} shows the decomposition of \emph{on-off keying} (OOK), \ie, $X \in \{0,1\}$, and \emph{phase-shift keying} (PSK) with $M=16$ phase levels.

As the input phase carries no information with OOK, the phase term and the mixed term~II are zero.
The amplitude term yields the amount of information available when only the signal amplitude is received and processed.
An example for such a system is the direct-detection receiver used in optical communication systems, where the photodiode responds to the incident light power~\cite[Ch.~4]{Agrawal}.
Receivers that have access to the full signal (amplitude and phase) can extract additional information about the input amplitude from the output phase.
This information gain is reflected in the mixed term~I (dotted line).
In the optical communications example, this gain can be obtained by upgrading an optical OOK system from direct to coherent detection.
At SNRs larger than 10~dB, all the information is contained in the received amplitude, so that receiving the signal phase does not yield any additional information.

For the PSK input, the amplitude term and the mixed term~I are zero.
A phase-only receiver captures most of the available information; the (rather small) gain that is obtained from additional amplitude reception at SNRs below 10 dB is captured by the mixed term~II (dash-dotted line).

\subsubsection{Combined ASK-PSK modulation}
\begin{figure}[!htb]
\centering
  \includegraphics[width=0.25\textwidth]{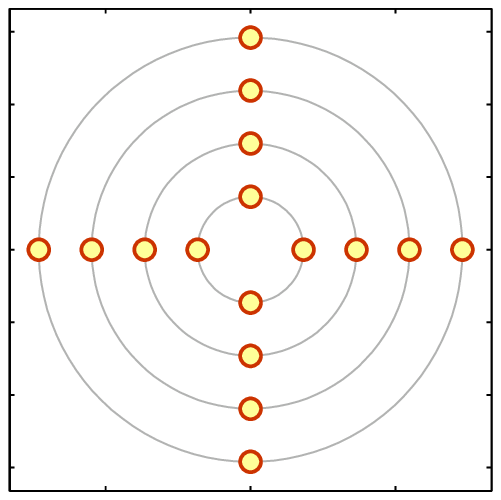}
  \includegraphics[width=0.25\textwidth]{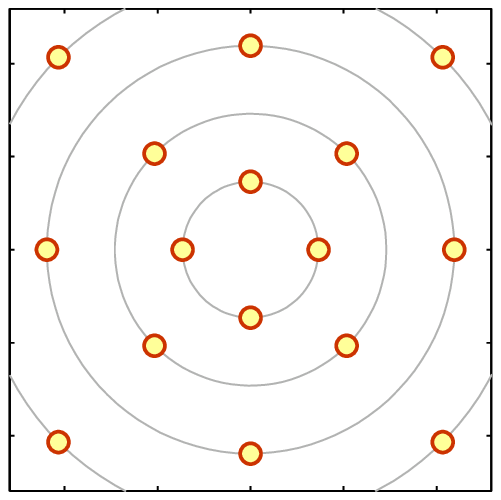}
  
  \includegraphics[width=0.25\textwidth]{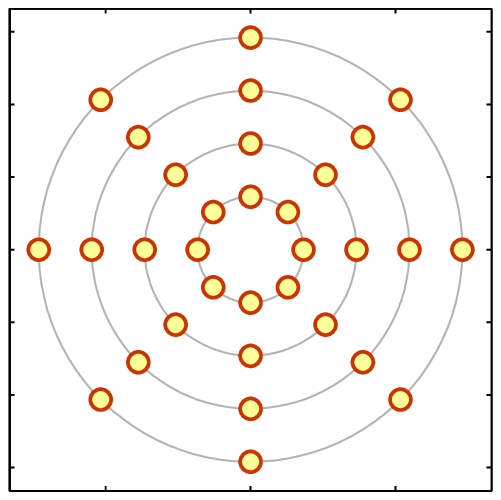}
  \includegraphics[width=0.25\textwidth]{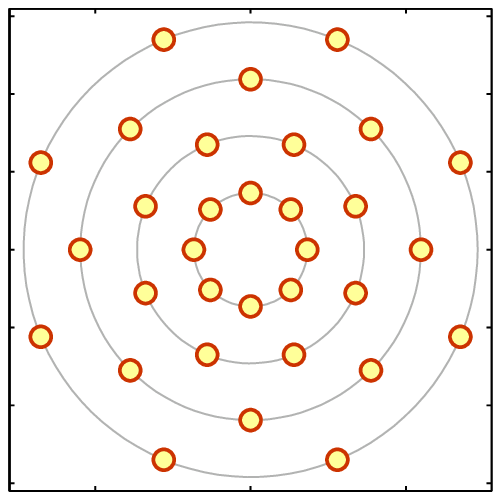}

  \includegraphics[width=0.25\textwidth]{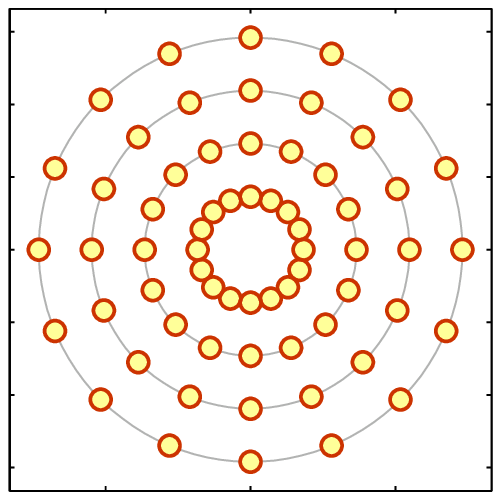}
  \includegraphics[width=0.25\textwidth]{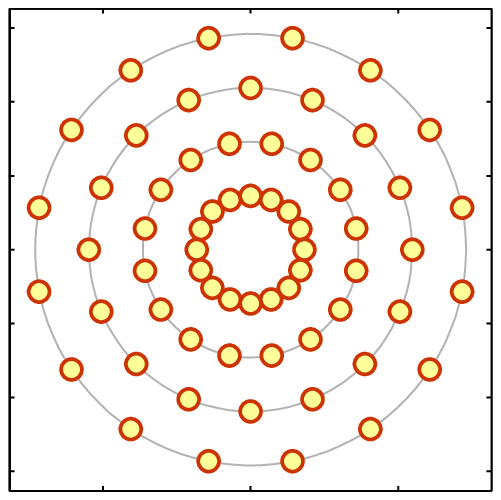}
\caption{Combined ASK/PSK signal constellations with 4 phase levels (top), 8 phase levels (center) and 16 phase levels (bottom) without (left) and with (right) phase shift. The number of amplitude levels is always 4.}\label{fig:constellations_ASKPSK}
\end{figure}

The simultaneous digital modulation of both amplitude and phase was first proposed in 1960~\cite{Cahn1960}.
Examples for this type of constellation, which later became known as the \emph{Type I} or \emph{star-QAM} constellation, are shown in Fig.~\ref{fig:constellations_ASKPSK}.
The constellations depicted in the left column are combinations of 4 amplitude levels and 4, 8, and 16 phase levels, respectively.
The constellations shown in the right column are modifications of these ASK/PSK schemes, where an additional phase offset is introduced between adjacent amplitude levels, thus increasing the minimum distance between neighboring constellation points.

\begin{figure}[!htb]
\centering
  \psfrag{C}[][]{\footnotesize Mutual information in bits per symbol}
  \psfrag{SNR}[][]{\footnotesize SNR in dB}
  \psfrag{Legend1xxxxxxxxx}{\scriptsize $I(X;Y)$}
  \psfrag{Legend2xxxxxxxxx}{\scriptsize $I(\sabs{X};\sabs{Y})$}
  \psfrag{Legend3xxxxxxxxx}{\scriptsize $I(\sang{X}; \sang{Y} | \sabs{X})$}
  \psfrag{Legend4xxxxxxxxx}{\scriptsize $I(\sabs{X}; \sang{Y} | \sabs{Y})$}
  \psfrag{Legend5xxxxxxxxx}{\scriptsize $I(\sang{X}; \sabs{Y} | \sabs{X},\sang{Y})$}
  \psfrag{4r}{\small 4-ASK/4-PSK}
  \psfrag{8r}{\small 4-ASK/8-PSK}
  \psfrag{16r}{\small 4-ASK/16-PSK}
  \psfrag{offset}{\small offset}
  \includegraphics[width=0.48\textwidth]{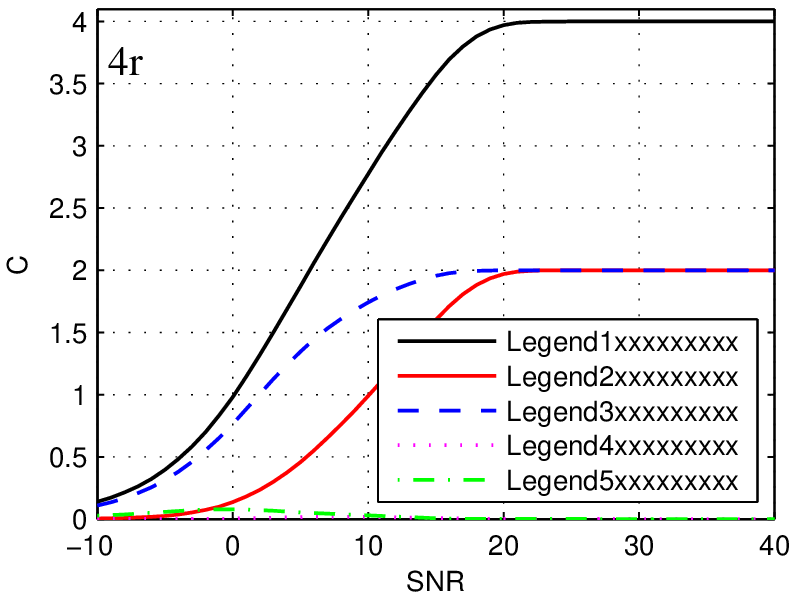}
  \includegraphics[width=0.48\textwidth]{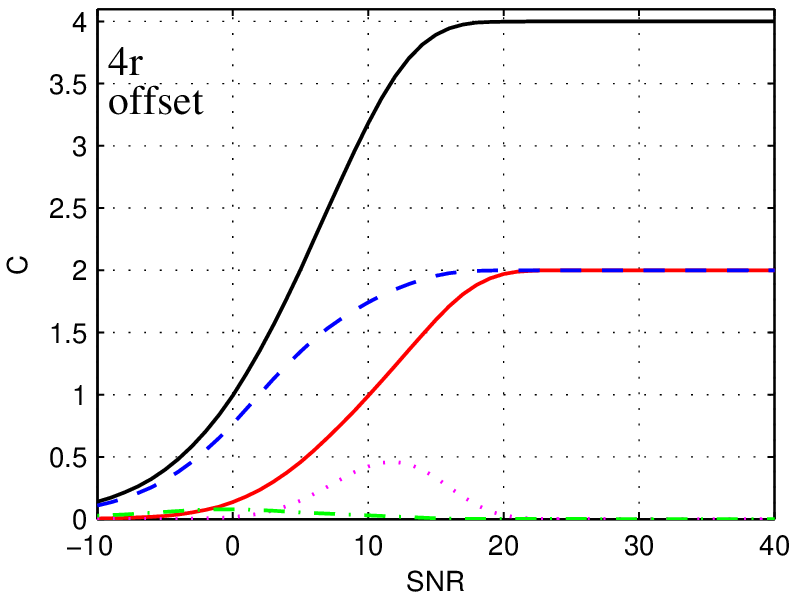}
  \includegraphics[width=0.48\textwidth]{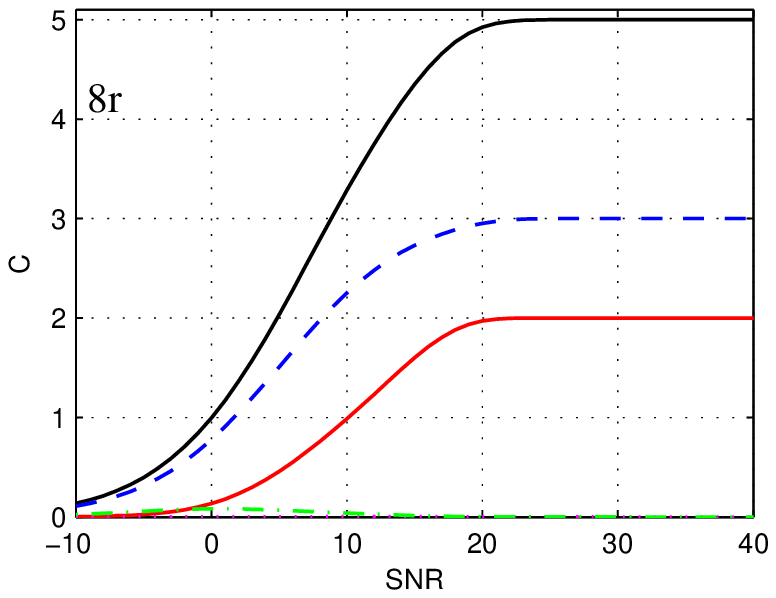}
  \includegraphics[width=0.48\textwidth]{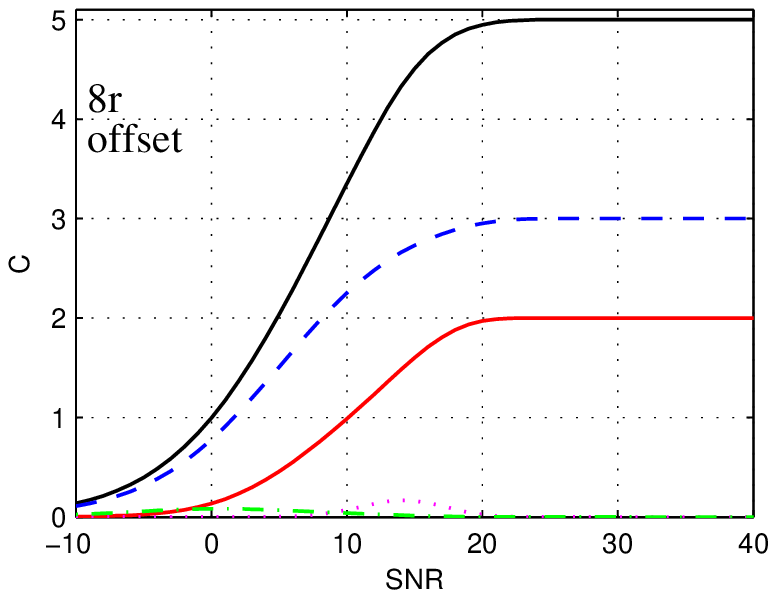}
  \includegraphics[width=0.48\textwidth]{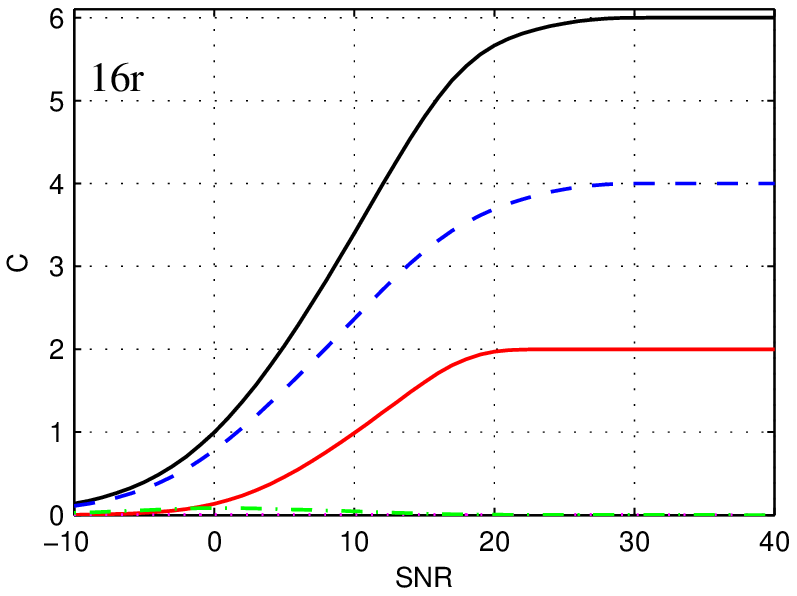}
  \includegraphics[width=0.48\textwidth]{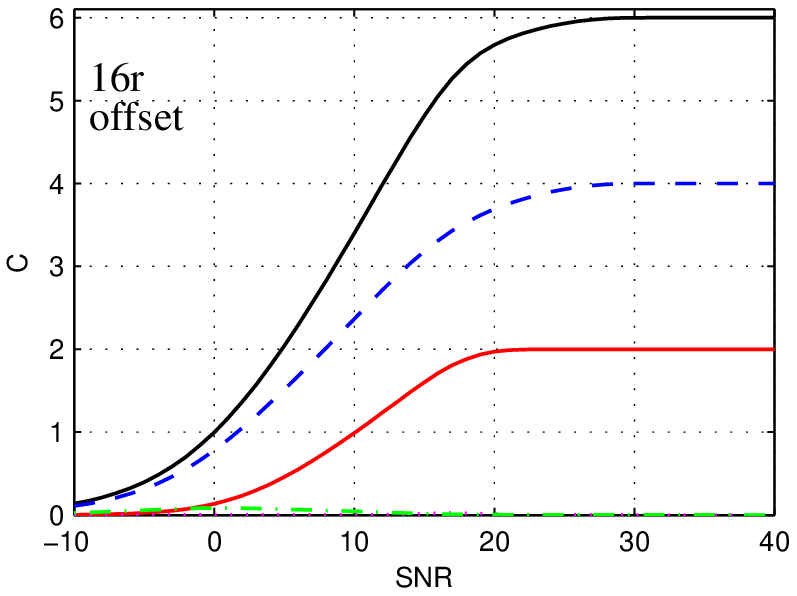}
\caption{Polar decomposition of mutual information for ASK/PSK constellations depicted in Fig.~\ref{fig:constellations_ASKPSK} without phase offset (left) and with phase offset (right).}\label{fig:decomposition_ASKPSK}
\end{figure}

The decomposition results shown in Fig.~\ref{fig:decomposition_ASKPSK} illustrate the capacity gain obtained from the phase offset in the right-column constellations of Fig.~\ref{fig:constellations_ASKPSK}.
As the joint amplitude PDF $p(\sabs{x},\sabs{y})$ remains unaffected by the phase offset, the amplitude term (red line) is equal for both constellations (compare plots on the left and on the right side of Fig.~\ref{fig:decomposition_ASKPSK}).
Similarly, the conditional joint phase PDF $p(\sang{x},\sang{y}|\sabs{x})$ only experiences a constant shift (along the $\sang{x}$-axis) for amplitude levels with phase offset, which does not change the decomposition phase term (blue line).
The capacity gain achieved by the phase offset is reflected in the increase of the mixed term~I (magenta line; \cf top left and top right plots); this gain decreases with increasing number of phase levels (top to bottom).

By letting the number of phase levels go to infinity, the constellation turns into continuous concentric rings and the mixed term~I tends towards zero.
Such modulation schemes with a discrete number of amplitude levels and continuous phase angles (so-called \emph{ring modulation}) were used in an extensive numerical study to estimate the capacity of nonlinear fiber-optic channels~\cite{Essiambre2010}.
As for the constellations discussed above (and most other constellations), the mixed term~II (green line) is negligibly small but non-zero for ASK/PSK constellations, too.

\subsubsection{QAM}
\begin{figure}[!htb]
\centering
  \psfrag{C}[][]{\footnotesize Mutual information in bits per symbol}
  \psfrag{SNR}[][]{\footnotesize SNR in dB}
  \psfrag{Legend1xxxxxxxxx}{\scriptsize $I(X;Y)$}
  \psfrag{Legend2xxxxxxxxx}{\scriptsize $I(\sabs{X};\sabs{Y})$}
  \psfrag{Legend3xxxxxxxxx}{\scriptsize $I(\sang{X}; \sang{Y} | \sabs{X})$}
  \psfrag{Legend4xxxxxxxxx}{\scriptsize $I(\sabs{X}; \sang{Y} | \sabs{Y})$}
  \psfrag{Legend5xxxxxxxxx}{\scriptsize $I(\sang{X}; \sabs{Y} | \sabs{X},\sang{Y})$}
  \psfrag{4QAM}{\small 4-QAM}
  \psfrag{16QAM}{\small 16-QAM}
  \psfrag{64QAM}{\small 64-QAM}
  \psfrag{256QAM}{\small 256-QAM}
  \psfrag{512QAM}{\small 512-QAM}
  \psfrag{1024QAM}{\small 1024-QAM}
  \includegraphics[width=0.48\textwidth]{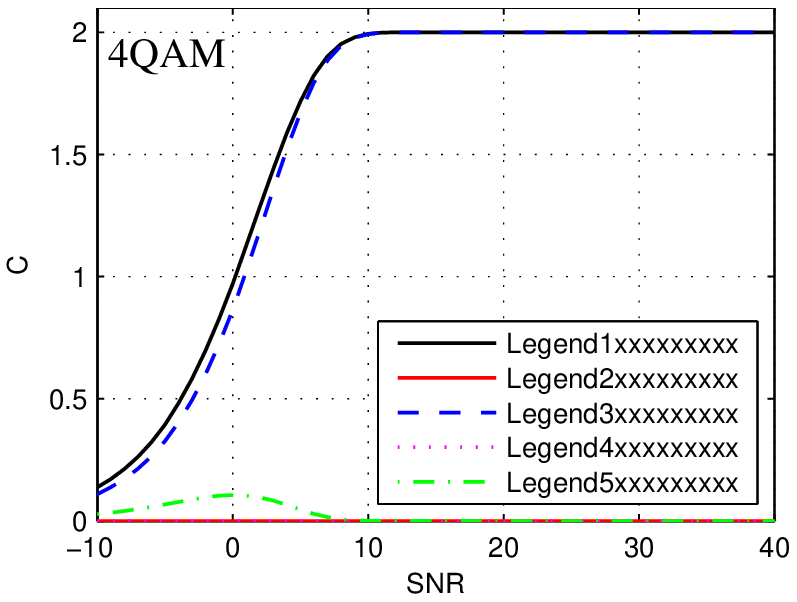}
  \includegraphics[width=0.48\textwidth]{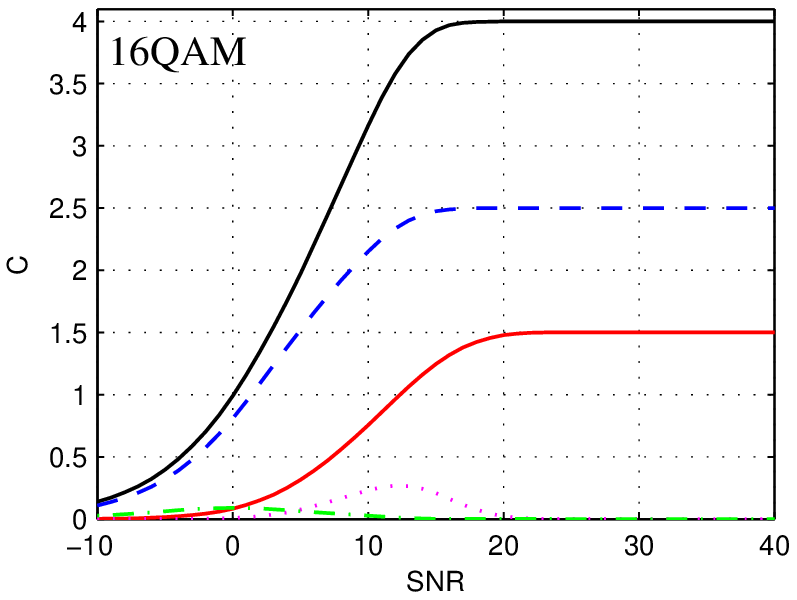}
  \includegraphics[width=0.48\textwidth]{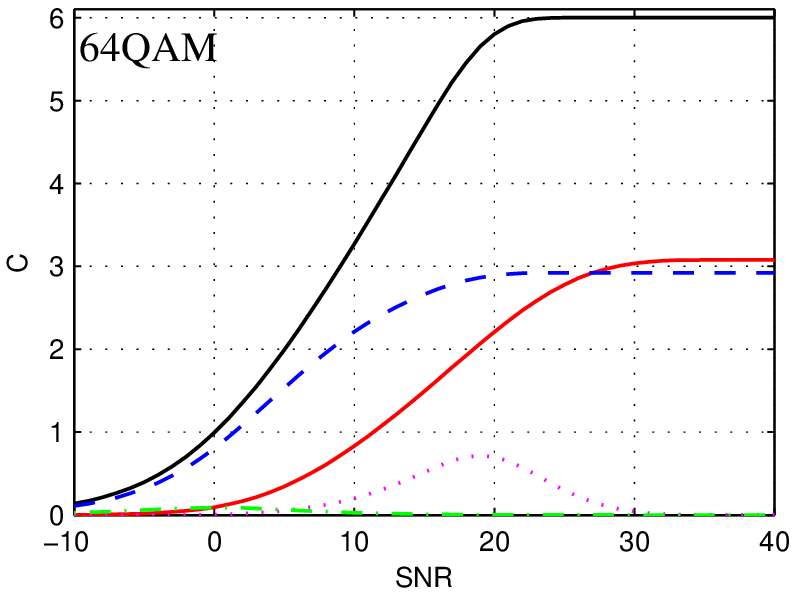}
  \includegraphics[width=0.48\textwidth]{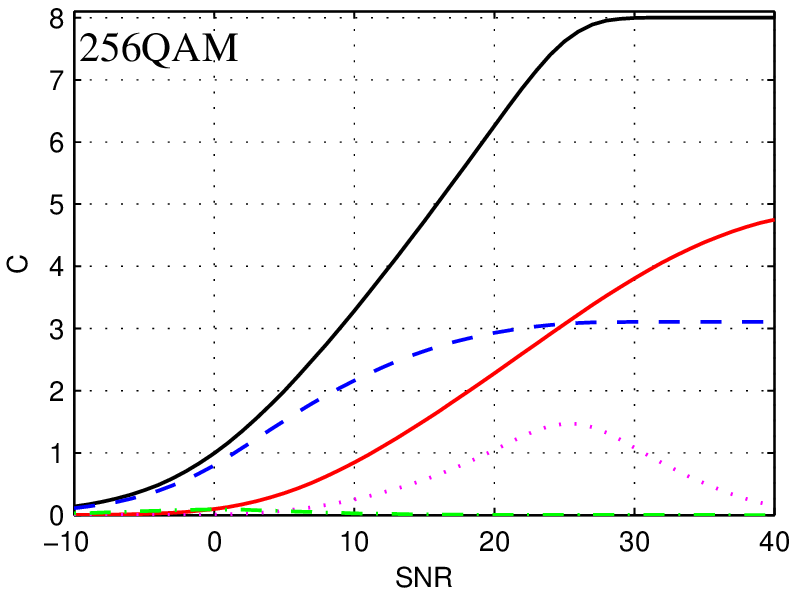}
  \includegraphics[width=0.48\textwidth]{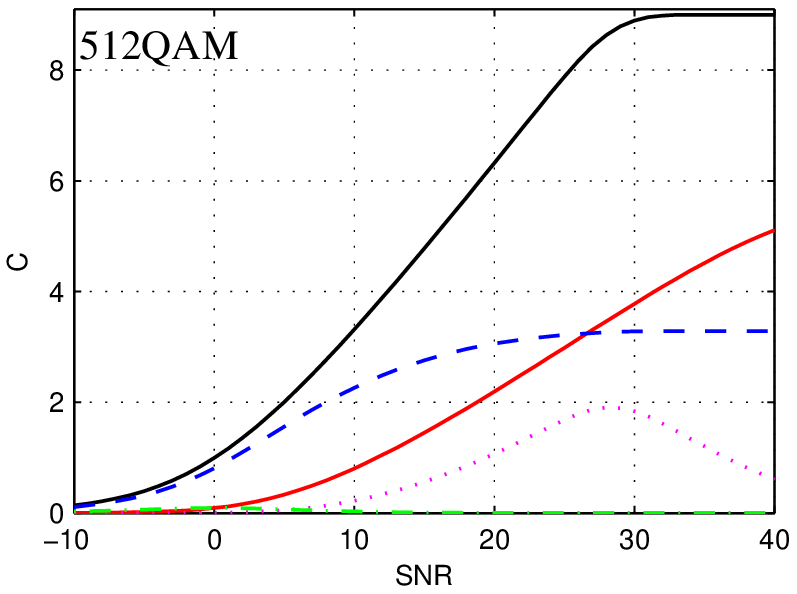}
  \includegraphics[width=0.48\textwidth]{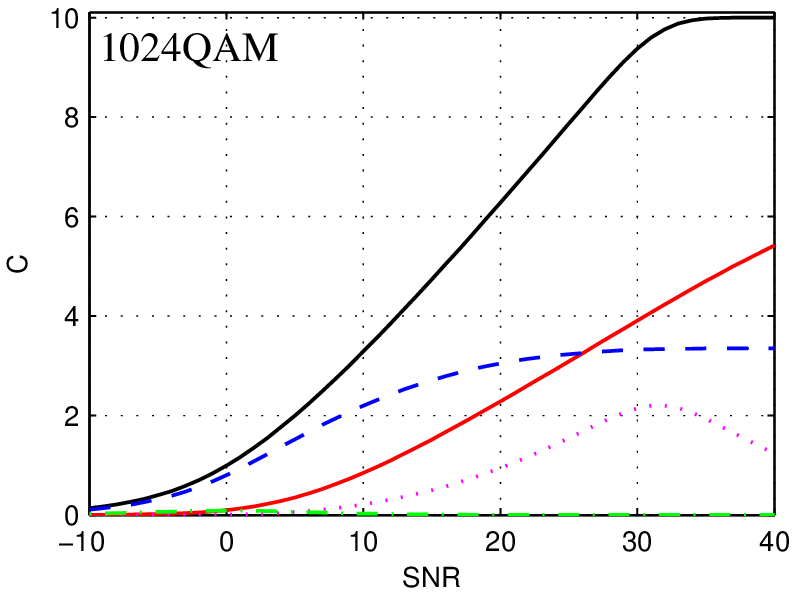}
\caption{Polar decomposition of mutual information for $M$-QAM constellations with (from top left to bottom right) $M=4,16,64,256,512,1024$.}\label{fig:decomposition_QAM}
\end{figure}

The polar decomposition results for $M$-QAM constellations with $M=4, 16, 64, 256, 512,$ $1024$ are shown in Fig.~\ref{fig:decomposition_QAM}.
It can be seen that the amplitude and phase terms saturate at $H(\sabs{X})$ and $H(\sang{X}|\sabs{X})$, respectively.
For instance, 16-QAM has three distinct amplitude levels with four or eight distinct phase levels each, so the decomposition terms tend towards
\begin{equation}
H(\sabs{X}) = -\frac{4}{16} \log_2 \left( \frac{4}{16} \right) - \frac{8}{16} \log_2 \left( \frac{8}{16} \right) - \frac{4}{16} \log_2 \left( \frac{4}{16} \right) = 1.5 \textrm{~bits}
\label{eq:16QAM_entropy_abs}
\end{equation}
and
\begin{equation}
H(\sang{X}|\sabs{X}) = \frac{1}{4} \log_2 4 + \frac{1}{2} \log_2 8 + \frac{1}{4} \log_2 4 = 2.5 \textrm{~bits}.
\label{eq:16QAM_entropy_ang}
\end{equation}

Among the considered QAM constellations, 4-QAM is a special case in the sense that its mixed term~I is zero; being a PSK format, its decomposition resembles that of 16-PSK depicted in Fig.~\ref{fig:decomposition_OOK_and_PSK}.
For $M>4$, QAM constellations exhibit a significant mixed term~I, so that in the analysis of this modulation scheme, the mutual information may not be approximated by the sum of the amplitude and phase terms only.
Again, mixed term~II is non-zero but negligibly small for all QAM constellations.

\clearpage

\section{Partially Coherent Channels}\label{sec:partially_coherent}
In the preceding section, the transmitted phase (and, of course, the amplitude, too) was corrupted by AWGN.
If the signal is impaired by phase noise (in addition to AWGN), the channel is only partially able to convey phase information even in the absence of AWGN.
Such channels are called%
\footnote{The term \emph{partially coherent} was introduced to communications engineering by A.~Viterbi in 1965~\cite{Viterbi1965}. Viterbi possibly adopted the term from physical optics, where it characterizes the temporal or spatial correlation of electrical fields that are neither \emph{coherent} (fully correlated) nor \emph{incoherent} (uncorrelated)~\cite[Ch.~X]{Born_Wolf}. In communication and information theory, the term \emph{noncoherent} (rather than \emph{incoherent}) is used to refer to channels that are entirely unable to transmit any phase information.}
\emph{partially coherent}~\cite{Katz2004}.
Partially coherent channels can be described in continuous-time form by
\begin{equation}
Y(t) = X(t) \cdot e^{\jm\Theta(t)} + N(t),
\label{eq:partially_coherent_channel_continuous_time}
\end{equation}
where $N(t)$ is a complex-valued AWGN process with variance $2\sigma_{\rm n}^2$ and $\Theta(t)$ models the phase noise process.
We can differentiate various types of phase noise appearing in communication systems:
\begin{itemize}
	\item The carrier itself as well as the local oscillator used for demodulation have random noise fluctuations. This type of phase noise occurs in lightwave communication systems where the laser phase performs a random walk (\eg, Brownian motion). The nonzero laser linewidth can broaden the signal spectrum so that spectrally sensitive operations (filtering, sampling) require special attention. References on laser phase noise and related system aspects include \cite{Salz1985, Foschini1988, Foschini1989, Shamai1989, Dallal1991}	and many references therein.
	\item Another type of correlated phase noise emerges when the carrier phase is imperfectly tracked at the receiver (\eg, in a phase-locked loop \cite{Foschini1973,Katz2004}). In this case, samples from the the phase noise process $\Theta(t)$ are usually assumed to have a von Mises (Tikhonov) distribution (\ref{eq:vonMises}).
	\item Uncorrelated (white) phase noise can be used to model the nonlinear effect of cross-phase modulation (XPM) \cite[Ch.~7]{AgrawalNL} in multi-channel fiber-optic communication systems. In this case, the phase noise samples follow a wrapped Gaussian distribution (\ref{eq:wrapped_Gaussian}) as explained in Sec.~\ref{subsec:capacity_optics}.
	\item Signal-dependent phase noise is also found in fiber-optic communication systems, where it is produced by the nonlinear effect of self-phase modulation (SPM) \cite[Ch.~5]{Ho_PM_systems_2005}. SPM induces a phase shift that is proportional to the instantaneous power of the propagating optical wave (including signal and noise) \cite[Ch.~4]{AgrawalNL}.
\end{itemize}

In general, all types of phase noise are capable of broadening the spectrum of the transmitted signal $X(t)$.
This spectral broadening is the major obstacle in transforming (\ref{eq:partially_coherent_channel_continuous_time}) into a discrete-time channel model.
Filtering (and sampling) a signal whose spectrum is broadened by phase noise can result (1) in signal distortions and energy loss when the filter is narrow \cite{Foschini1988, Dallal1991} and (2) in an increased captured noise power when the filter bandwidth is wide (see \cite{Dallal1991} and references therein).
These effects can be neglected when the spectral broadening is moderate, which is the case for strongly correlated phase noise processes.
Filtering and sampling at the symbol rate is then possible and leads to discrete-time channel models that have independent and identically distributed (\iid) signal and noise samples, but correlated phase noise samples (see, \eg, \cite{Peleg2000}).
To obtain a discrete-time channel model with uncorrelated phase noise samples, the presence of an ideal interleaver and de-interleaver can be assumed (\eg, \cite{Katz2004}).
It is then possible to transform (\ref{eq:partially_coherent_channel_continuous_time}) into the discrete-time form
\begin{equation}
Y = X \cdot e^{\jm\Theta} + N,
\label{eq:partially_coherent_channel_discrete_time}
\end{equation}
in which the phase noise time samples $\Theta_i$ are modeled as \iid and statistically independent of $X$.
In Sec.~\ref{subsec:spectral_loss}, we instead discuss partially coherent channels with \emph{white} phase noise.
Discretization of such channels by means of filtering and sampling at the symbol rate is possible, but does not lead to (\ref{eq:partially_coherent_channel_discrete_time}).
Instead, the discrete-time channel model must be modified to account for an effect we call \emph{spectral loss}.

Before continuing with information rates, we remark that since the phase angle of AWGN is uniformly distributed, the order in which phase noise and AWGN act on the transmitted signal is irrelevant:
\begin{align}
Y &= (X+N) \cdot e^{\jm\Theta} \nonumber \\
 &= X \cdot e^{\jm\Theta} + N \cdot e^{\jm\Theta} \nonumber \\
 &= X \cdot e^{\jm\Theta} + N',
\label{eq:partially_coherent_channel2}
\end{align}
where $N' \sim \normalc{0}{2\sigma_{\rm n}^2}$ has the same distribution as $N$.

The circular PDF $p(\sang{Y})$ can be obtained by circular convolution~\cite{Gray1995} of (\ref{eq:phase_PDF_AWGN}) with $p(\Theta)$.
In numerical experiments, it is usually more efficient to multiply the PDFs' discrete Fourier transforms (DFT) and perform an inverse DFT (IDFT) to obtain the final result~\cite{Jeruchim2000}.
In particular, when the phase noise has a wrapped Gaussian distribution, the DFT of (\ref{eq:phase_PDF_AWGN}) can be multiplied with the DFT of the ``unwrapped'' Gaussian (which is again Gaussian).
The following IDFT will implicitly ``wrap'' the resulting PDF so that it maintains its periodicity with $2\pi$.

In the following discussion of partially coherent channels, the term \emph{SNR} refers to the power ratio of signal and \emph{additive} noise (\ref{eq:SNR}).

\subsection{Input Optimization and Information Rate Calculation}
The capacity-achieving input distribution for the partially coherent channel (\ref{eq:partially_coherent_channel_discrete_time}) is not Gaussian~\cite{Hou2002}, but it is circularly symmetric~\cite{Hou2002}, \ie, uniform in phase, and has discrete amplitude levels~\cite{Hou2003,Katz2004}.
In other words, the capacity-achieving input distribution for the partially coherent channel consists of a number of continuous rings; the number, radii and probabilities of these rings are subject to optimization.
Interestingly, the shaping gain that can be achieved by using non-equiprobable input symbols rather than a uniform square-area or circular-area distribution is significantly larger than the maximum shaping gain of 1.53~dB for the AWGN channel~\cite{Hou2002}.
Therefore, the optimization of signal sets for the partially coherent channel may be more rewarding than for the AWGN channel.

The polar decomposition is useful for the analysis of partially coherent channels when both mixed terms are small or can be neglected.
As mentioned in Section~\ref{sec:AWGN}, this is the case for AWGN channels with Gaussian or ring inputs.
As the amplitude term $I(\sabs{X};\sabs{Y})$ is not affected by phase noise, it suffices to re-calculate the phase term in the presence of phase noise.
The conditional phase PDF $p(\sang{y}|\sabs{x},\sang{x}\!=\!0)$ is obtained numerically or, where possible, analytically from a circular convolution~\cite{Gray1995} of (\ref{eq:phase_PDF_AWGN}) with the phase noise PDF, usually (\ref{eq:wrapped_Gaussian}) or (\ref{eq:vonMises}).

\begin{figure}[!htb]
\centering
  \psfrag{C}[][]{\footnotesize Mutual information in bits per symbol}
  \psfrag{SNR}[][]{\footnotesize SNR in dB}
  \psfrag{s00}{$\sigma=0$}
  \psfrag{s05}{$\sigma=0.5$}
  \psfrag{s1}{$\sigma=1$}
  \psfrag{s2}{$\sigma=2$}
  \psfrag{Legend1xxxxx}{\scriptsize $I(X;Y)$}
  \psfrag{Legend2xxxxx}{\scriptsize $I(\sabs{X};\sabs{Y})$}
  \psfrag{Legend3xxxxx}{\scriptsize $I(\sang{X}; \sang{Y} | \sabs{X})$}
  \includegraphics[width=0.48\textwidth]{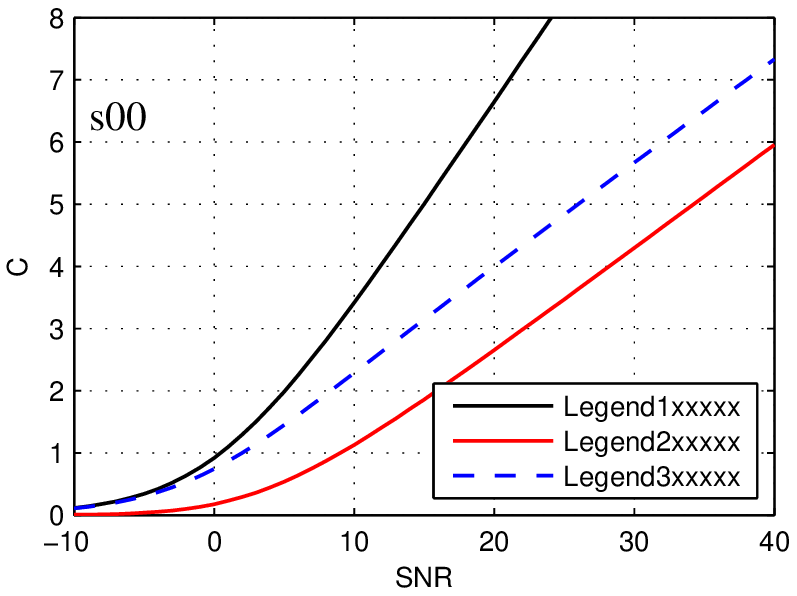}
  \includegraphics[width=0.48\textwidth]{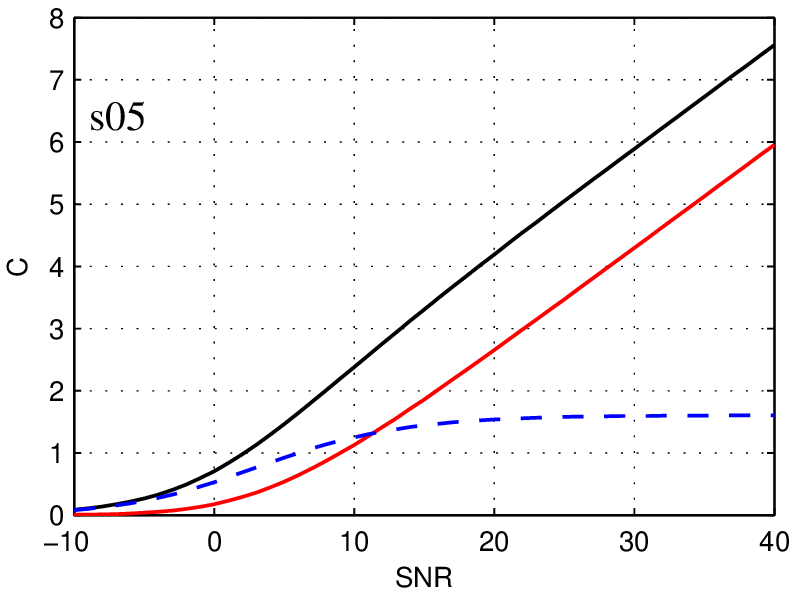}
  \includegraphics[width=0.48\textwidth]{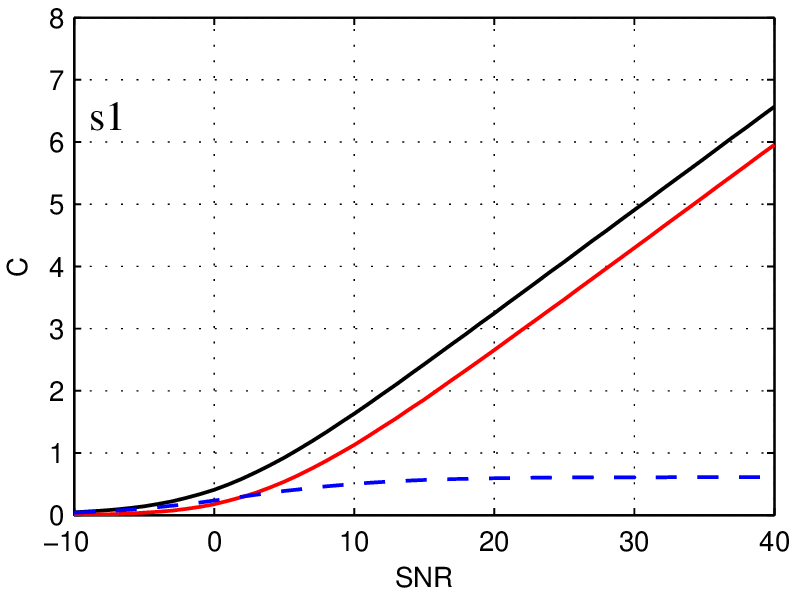}
  \includegraphics[width=0.48\textwidth]{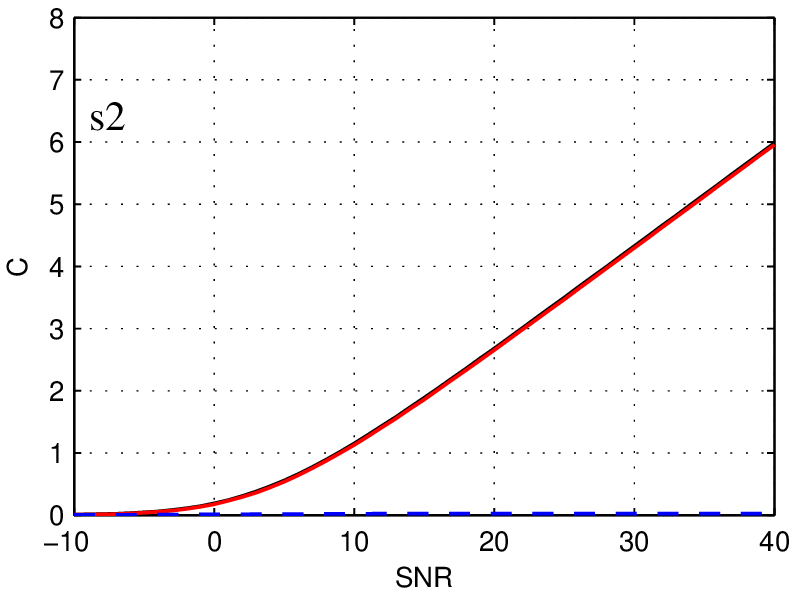}
\caption{Polar decomposition of mutual information for an AWGN channel with Gaussian input with additional phase noise ($\sigma \!=\! 0, 0.5, 1, 2$). The mixed term~II is negligible.}\label{fig:decomposition_partially_coherent}
\end{figure}

Fig.~\ref{fig:decomposition_partially_coherent} shows the decomposition results for the AWGN channel with Gaussian input with additional phase noise.
The phase noise has a wrapped Gaussian distribution with parameter $\sigma$ as shown in Fig.~\ref{fig:wG_vM_PDFs}.%
\footnote{We remind the reader that $\sigma_{\rm n}^2$ denotes the AWGN's variance per dimension, whereas $\sigma$ is the parameter of the wrapped Gaussian distribution. Note that this distribution's circular variance is given by (\ref{eq:wrapped_Gaussian_circularvariance}); it is not equal to $\sigma^2$.}
The phase noise parameter values are $\sigma \!=\! 0, 0.5, 1, 2$.
For large $\sigma$, the circular variance goes to one and the wrapped Gaussian distribution becomes uniform.
In this case, no information can be transmitted in the signal phase and the phase term tends to zero.
An interesting observation can be made when $\sigma$ is small (but nonzero).
In this case, the phase term increases with increasing SNR, but tends towards a constant value asymptotically.
When the phase term nearly reaches this asymptote, the contribution of the phase term to $I(X;Y)$ gets small compared to that of the amplitude term (which rises logarithmically with the SNR, \cf (\ref{eq:AWGN_amplitude_term_asymptotic})).
This statement is valid for any (arbitrarily low) phase noise variance.
Fig.~\ref{fig:decomposition_partially_coherent} shows the amplitude term and the phase terms for $\sigma \!=\! 0, 0.5, 1, 2$ and the respective total capacities.
The (very small) contribution of the mixed term~II was neglected.

\subsection{Noncoherent Channels}\label{subsec:noncoherent}
The \emph{noncoherent} channel is the limiting case of the partially coherent channel (\ref{eq:partially_coherent_channel_discrete_time}) when $\Theta$ is distributed uniformly in $[-\pi, \pi)$.
As the phase is completely randomized, the output phase $\sang{y}$ carries no information and the phase term and the mixed term~I of the polar decomposition are zero.
The fact that the mixed term~II is zero, too, is a consequence of $p(\sabs{y}|\sabs{x},\sang{y})=p(\sabs{y}|x,\sang{y})$.
The only information that can be transmitted over the noncoherent channel is, therefore, represented by the amplitude term $I(\sabs{X};\sabs{Y})$.
An example for a noncoherent channel is the previously mentioned optical direct-detection (DD) receiver, which can be modeled by $Y \!=\! |X \!+\! N|^2$.

A related but different situation occurs for channels that obey $Y \!=\! |X|^2 \!+\! N$.
Channels of this kind are found in a variety of optical communication scenarios, with different statistics for $N$.
For example, in thermal-noise limited DD receivers $N$ is a Gaussian process, but other noise statistics can be found for channels limited by (multiplied) shot noise or by large amounts of optical background noise, both in fiber and free-space optical communications.
For a discussion of optical intensity channels with AWGN, see, \eg, \cite{Hranilovic2004}.
While the phase term and the mixed term~II are zero in this case as for the noncoherent and DD channels, the mixed term~I can be larger than zero.
Similarly, when the channel input is constrained to real-valued amplitude modulation, \ie, when the channel model is $Y \!=\! X+N, \,\, \mcal{X} \!=\! [0, \infty),$ the mixed term~I can be larger than zero.
The decomposition of the AWGN channel with OOK modulation discussed in Section~\ref{subsec:AWGN_discrete} is an example.

The conditional PDF $p(\sabs{y}|\sabs{x})$ of the noncoherent channel is Ricean (\ref{eq:Ricean_distribution}), so the mutual information $I(\sabs{X};\sabs{Y})$ is calculated along the lines of the amplitude term calculation in Section~\ref{subsec:AWGN_Gaussian}.
The difficulty in finding the capacity of the noncoherent channel lies in finding the optimum input distribution $p(\sabs{x})$.
Similar to the partially coherent channel, it is known that the optimum input distribution $p(x)$ is not Gaussian~\cite{Colavolpe2001}, \ie, the optimum $p(\sabs{x})$ is not a Rayleigh distribution (\ref{eq:Rayleigh_distribution}).
Rather, the capacity-achieving input is discrete~\cite{Katz2004}.
By numerical optimization, Ho found an optimum input (for the optical DD channel) that has a discrete probability mass at $\sabs{x}=0$ and a continuous exponential profile at $\sabs{x}>0$~\cite{Ho2005}.
At low SNRs, this distribution collapses to two discrete points at $\sabs{x}=0$ and at $\sabs{x}>0$, \ie, OOK, confirming a result reported in~\cite{Katz2004}.

An analytical approximation to the noncoherent channel's capacity is available in the limit of large SNRs.
In this case, the Ricean distribution $p(\sabs{y}|\sabs{x})$ can be approximated by a Gaussian, and the capacity-achieving input distribution is a \emph{positive normal} or \emph{half-Gaussian} distribution~\cite{Blachman1953}
\begin{equation}
p(\sabs{x}) = \left\{
\begin{array}{ll}
	\sqrt{\frac{2}{\pi P_{\rm s}}} \cdot \exp\left( -\frac{\sabs{x}^2}{2 P_{\rm s}} \right), & x \geq 0, \\
	0, & x < 0.
\end{array}
\right.
\label{eq:half-Gaussian_PDF}
\end{equation}
In a derivation analogous to that of (\ref{eq:AWGN_amplitude_term_asymptotic}), the capacity is found to be~\cite{Blachman1953}
\begin{equation}
I(\sabs{x};\sabs{y}) \approx \frac{1}{2} \log_2 \left( \frac{P_{\rm s}}{2\sigma_{\rm n}^2} \right) - \frac{1}{2}, \quad P_{\rm s} \gg 2\sigma_{\rm n}^2,
\label{eq:capacity_noncoherent_asymptotic}
\end{equation}
which is $(\log_2\pi - (1+\gamma)/(\ln 2) + 1)/2 \approx 0.19$ bits higher than the mutual information (\ref{eq:AWGN_amplitude_term_asymptotic}) that results from a Rayleigh-distributed input.
The same result was found in an analysis of optical DD systems~\cite{Mecozzi2001}.
Signal shaping methods for the optical DD are discussed in~\cite{Mao2008}.

\subsection{Spectral Loss Induced by White Phase Noise}\label{subsec:spectral_loss}
As discussed above, certain types of phase noise induce spectral broadening.
If the phase noise process $\Theta(t)$ is white, \ie, if it is temporally uncorrelated, a related but qualitatively different effect occurs which we call \emph{spectral loss}.

To describe this effect, we use the continuous-time channel model (\ref{eq:partially_coherent_channel_continuous_time}).
We derive the power spectral density (PSD) $\iPhi_{Y}(f)$ of $Y(t)$, assuming that $X(t)$ and $N_{\rm PN}(t)=e^{\jm\Theta(t)}$ are stationary, ergodic, and statistically independent random processes.
The autocorrelation function (ACF) $\vphi_{Y}(\tau)$ of $Y(t)$ is~\cite{OhmLueke}
\begin{align}
\vphi_{Y}(\tau) &= \expectation{}{X(t) \cdot e^{\jm\Theta(t)} \cdot X^\ast(t+\tau) \cdot e^{-\jm\Theta(t+\tau)}} + \expectation{}{N(t) \cdot N^\ast(t+\tau)} \nonumber \\
 &= \expectation{}{X(t) \cdot X^\ast(t+\tau)} \cdot
		\expectation{}{e^{\jm\Theta(t)} \cdot e^{-\jm\Theta(t+\tau)}} + \vphi_{N}(\tau) \nonumber \\
 &= \vphi_{X}(\tau) \cdot \vphi_{N_{\rm PN}}(\tau) + \vphi_{N}(\tau),
\label{eq:partially_coherent_channel_time_continuous_ACF}
\end{align}
where $\expectation{}{.}$ denotes the ensemble average.
In calculating the ACF $\vphi_{N_{\rm PN}}(\tau)$ of $N_{\rm PN}(t)$, we assume for simplicity that the phase noise follows a wrapped Gaussian distribution (\ref{eq:wrapped_Gaussian}) with parameter $\sigma$.
Since $\Theta(t)$ and $\Theta(t+\tau)$ are independent samples of a Gaussian random process, their sum or difference $\Theta'(t) \!=\! \Theta(t) \pm \Theta(t+\tau)$ satisfies $\Theta'(t) \sim \normalr{0}{2\sigma^2}$ for $\tau \neq 0$.
The autocorrelation function $\vphi_{N_{\rm PN}}(\tau)$ of the phase noise process $N_{\rm PN}(t)$ is
\begin{align}
\vphi_{N_{\rm PN}}(\tau) &= \expectation{}{e^{\jm\Theta(t)} \cdot e^{-\jm\Theta(t+\tau)}} \nonumber \\
 &= \left\{
		\begin{array}{ll}
			1, & \tau = 0, \\
			e^{-\sigma^2}, & \tau \neq 0,
		\end{array} \right.
\label{eq:phase_noise_ACF}
\end{align}
where the last result (for $\tau\neq 0$) is the resultant length (\ref{eq:wrapped_Gaussian_resultantlength}) of an ergodic (wrapped) Gaussian random variable $\Theta'$ with zero mean and variance $2\sigma^2$:
\begin{equation}
\expectation{}{e^{\jm(\Theta(t)-\Theta(t+\tau))}} =
\expectation{}{e^{\jm\Theta'}} = e^{-\sigma^2}, \quad \tau \neq 0.
\label{eq:phase_noise_ACF_calculation_detail}
\end{equation}

The piecewise defined ACF (\ref{eq:phase_noise_ACF}) can be written as
\begin{equation}
\vphi_{N_{\rm PN}}(\tau) = e^{-\sigma^2} +
		\lim_{B\to\infty} (1-e^{-\sigma^2}) \cdot \sinc{B \tau},
\label{eq:phase_noise_ACF2}
\end{equation}
where $\sinc{x} \!=\! \sin(\pi x)/(\pi x)$.
By the Wiener-Khinchin theorem~\cite{OhmLueke}, the PSD $\iPhi_{N_{\rm PN}}(f)$ of $N_{\rm PN}(t)$ is
\begin{align}
\iPhi_{N_{\rm PN}}(f) &= \Fourier{\vphi_{N_{\rm PN}}(\tau)} \nonumber \\
 &= e^{-\sigma^2} \delta(f) + \lim_{B\to\infty} (1-e^{-\sigma^2}) \cdot \frac{1}{B} \cdot \rect{\frac{f}{B}},
\label{eq:phase_noise_PSD}
\end{align}
where
\begin{equation}
\rect{f} = \left\{
	\begin{array}{ll}
		1, & |f| < \frac{1}{2}, \\
		\frac{1}{2}, & |f| = \frac{1}{2}, \\
		0, & |f| > \frac{1}{2}
	\end{array} \right\}	= \Fourier{\sinc{t}}
\label{eq:rect_function}
\end{equation}
is the \emph{rectangular function} \cite{Gray1995}.
Finally, the PSD $\iPhi_Y(f)$ of $Y(t)$ is calculated using (\ref{eq:partially_coherent_channel_time_continuous_ACF}) and (\ref{eq:phase_noise_PSD}) as
\begin{align}
\iPhi_{Y}(f) &= \iPhi_X(f) \star \iPhi_{N_{\rm PN}}(f) + \iPhi_{N}(f) \nonumber \\
 &= e^{-\sigma^2} \iPhi_{X}(f) + \lim_{B\to\infty} \iPhi_X(f) \star (1-e^{-\sigma^2}) \frac{\rect{f/B}}{B} + \iPhi_{N}(f).
\label{eq:partially_coherent_channel_time_continuous_PSD}
\end{align}
The $\star$ sign denotes convolution.
Equation (\ref{eq:partially_coherent_channel_time_continuous_PSD}) explains the spectral effect of phase noise: The original PSD $\iPhi_X(f)$ is preserved in shape, but attenuated by a factor $e^{-\sigma^2}$.
The remaining signal power (a fraction of $1-e^{-\sigma^2}$) is spread over the entire spectrum from $-\infty$ to $+\infty$ through convolution with a rectangular function whose width goes to infinity and whose height tends to zero.
Because the fraction of power that leaks outside the original spectrum has arbitrarily low power in any given finite band, it does not appear as spectral interference.
Hence, we call the effect \emph{spectral loss} (in contrast to \emph{spectral broadening}).

A remarkable feature of (\ref{eq:partially_coherent_channel_time_continuous_PSD}) is its simplicity: the original PSD $\iPhi_X(f)$ is not broadened.
We conclude from (\ref{eq:partially_coherent_channel_time_continuous_PSD}) that filtering (at the bandwidth of $\iPhi_X(f)$) the output of the partially coherent channel with white phase noise and sampling at the Nyquist rate produces a channel of the form
\begin{equation}
Y = X \cdot e^{-\frac{1}{2} \cdot \sigma^2} \cdot e^{\jm \Upsilon} + N,
\label{eq:partially_coherent_channel_discrete_time_spectralloss_plus_PN}
\end{equation}
where $\Upsilon$ is a random variable and the factor $1/2$ appears since (\ref{eq:partially_coherent_channel_discrete_time_spectralloss_plus_PN}) is expressed in terms of amplitudes.
In fact, numerical simulations (with large $B$) show that $\Upsilon$ approches $0$ as $B$ increases.
The resulting discrete-time model for our channel is
\begin{equation}
Y = X \cdot e^{-\frac{1}{2} \cdot \sigma^2} + N.
\label{eq:partially_coherent_channel_discrete_time_spectralloss}
\end{equation}
Eq.~(\ref{eq:partially_coherent_channel_discrete_time_spectralloss}) models an AWGN channel whose SNR (\ref{eq:SNR}) is attenuated by $e^{-\sigma^2}$, so this channel's capacity is
\begin{equation}
C = \log_2 \left( 1 + \frac{P_{\rm s} \cdot e^{-\sigma^2}}{2\sigma_{\rm n}^2} \right).
\label{eq:partially_coherent_channel_discrete_time_spectralloss_capacity}
\end{equation}
If the phase noise distribution is not (wrapped) Gaussian, the same calculation will lead to qualitatively similar results, with the value of the ACF (\ref{eq:phase_noise_ACF}) at $\tau \neq 0$ determining the spectral loss factor.

We remark that (\ref{eq:partially_coherent_channel_time_continuous_PSD}) has an important implication for numerical simulations of phase noise.
Due to the infinite spectral broadening of the power, the output signal $Y(t)$ has infinite bandwidth and is therefore necessarily undersampled in numerical simulations with finite bandwidth.
Therefore, the numerical simulation of phase noise will create aliasing inside and outside the original signal band through convolution of $\iPhi_X(f)$ with a rectangular function of finite width and nonzero height.
To keep this aliasing effect small in numerical simulations, it is necessary to oversample $X(t)$ by a sufficiently large factor and to filter the spurious out-of-band noise.

\subsection{Capacity of Nonlinear Fiber-optic Communication Channels}\label{subsec:capacity_optics}
Fiber-optic systems are one example for a channel that can be impaired by phase noise.
It is therefore tempting to apply the channel model with phase noise and spectral loss introduced above to estimate the channel capacity of fiber-optic systems in certain cases.
Such systems either transport a single channel or carry multiple channels via wavelength division multiplexing (WDM).
In general, capacity calculations for this channel are very difficult due to the medium nonlinearity, its interaction with linear channel effects, and the distributed nature of noise, nonlinearity and dispersion.
A method for estimating this channel's capacity from exhaustive numerical simulations is proposed in~\cite{Essiambre2010} where results for different physical scenarios are reported.
We first emphasize that the curves in \cite[Fig.~36]{Essiambre2010} are for a fixed AWGN variance, except for Curve (2) where the AWGN variance is set to zero.
Hence increasing SNR at fixed system length refers to increasing the transmit signal power.
At low signal power levels, the fiber channel is dominated by ASE (\emph{Amplified Spontaneous Emission}) noise from optical amplifiers and can be characterized as an AWGN channel.
With increasing signal power, distortions from nonlinear fiber effects increase faster than the SNR, bringing the channel capacity down to zero eventually.
\emph{Cross-phase modulation} (XPM) is identified as the most relevant effect for the channel capacity of WDM systems \cite[Fig.~36]{Essiambre2010}.
XPM causes a modulation of the signal phase in one WDM channel by the instantaneous power levels of all co-propagating channels~\cite[Ch.~7]{AgrawalNL}.
Single-channel systems have a higher capacity because of the absence of such \emph{inter-channel} nonlinearities \cite[Fig.~36, Curves (3) and (4)]{Essiambre2010}.
There, the fundamentally limiting effect involves the nonlinear interaction of signal and ASE noise.

Separate results for two special cases give insights into the origin of the capacity limitations \cite[Sec.~XI-E]{Essiambre2010}: (1) If XPM is suppressed (by transmitting one channel only), then the capacity starts decreasing at a much higher SNR than with WDM, see \cite[Fig.~36, Curves (3) and (4)]{Essiambre2010}. (2) In the ``unphysical'' case where ASE noise is absent (but all WDM channels are present), the capacity is still limited by XPM, see \cite[Fig.~36, Curve (2)]{Essiambre2010}.


In the following, we will concentrate on the single-channel case (with optical filtering) which is limited by nonlinear signal-noise interaction \cite[Fig.~36, Curve (3)]{Essiambre2010}.
In contrast to all other cases considered in \cite{Essiambre2010}, the capacity for the single-channel system setup decreases sharply with SNR.
To reproduce this curve with the channel model (\ref{eq:partially_coherent_channel_discrete_time_spectralloss}), we assume that the phase noise variance $\sigma^2$ in (\ref{eq:partially_coherent_channel_discrete_time_spectralloss}) scales quadratically with $P_{\rm s}$ (\ie, $\sigma^2 \!=\! c \cdot P_{\rm s}^2$, where $c$ is a constant).
Stated differently, we assume that the amplitude of the phase shift fluctuations induced by SPM scales linearly
with the signal power.

Using this model, a rapid capacity loss occurs (see (\ref{eq:partially_coherent_channel_discrete_time_spectralloss_capacity})) if the channel suffers from spectral loss.
More precisely, at high powers $P_{\rm s}$ (\ref{eq:partially_coherent_channel_discrete_time_spectralloss_capacity}) gives
\begin{equation}
C = \log_2 \left( 1 + \frac{P_{\rm s} \cdot e^{-cP_{\rm s}^2}}{2\sigma_{\rm n}^2} \right)
	\approx \frac{P_{\rm s} \cdot e^{-cP_{\rm s}^2}}{2\sigma_{\rm n}^2} \cdot \log_2(e).
\label{eq:partially_coherent_channel_discrete_time_spectralloss_capacity_2}
\end{equation}

\begin{figure}[!htb]
\centering
  \psfrag{C}[][]{Capacity in bits per symbol}
  \psfrag{SNR}[][]{SNR in dB}
  \psfrag{P}[][]{Avg.~optical power in dBm}
  \psfrag{Legend1xxxxxxxxx}{\small AWGN channel}
  \psfrag{Legend2xxxxxxxxx}{\small Channel (\ref{eq:partially_coherent_channel_discrete_time_spectralloss})}
  \psfrag{Legend3xxxxxxxxx}{\small Num.~results \cite{Essiambre2010}}
  \includegraphics[width=0.6\textwidth]{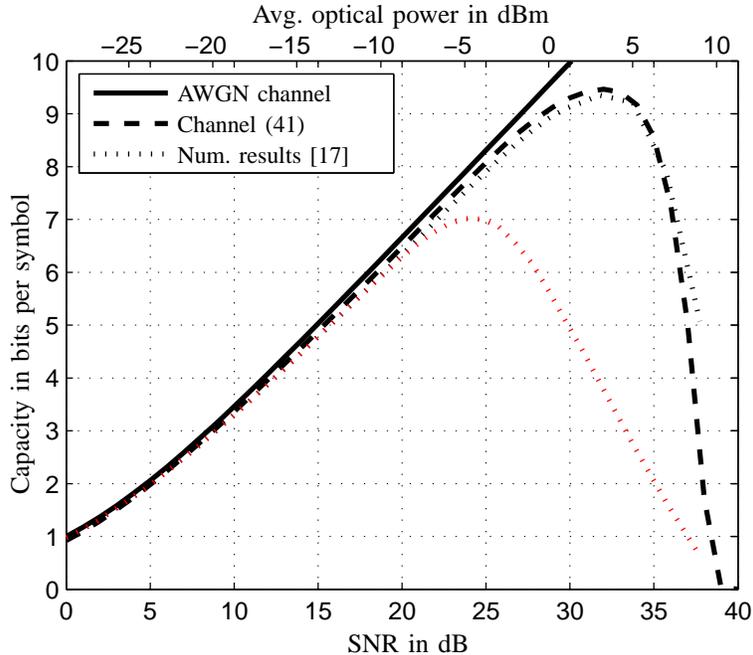}
\caption{Capacities of the fiber-optic channel (16-ring constellation input, single propagating channel) modeled as partially coherent channel (\ref{eq:partially_coherent_channel_discrete_time_spectralloss}) with wrapped Gaussian phase noise distribution with $\sigma^2 \!=\! c \cdot P_{\rm s}^2$. Dotted lines show numerical results from \cite[Fig.~36, Curves (1) and (3)]{Essiambre2010} (single channel (black) and WDM system (red)). Upper x-axis shows $P_{\rm s}$ in dBm, lower x-axis shows SNR in dB (\ref{eq:SNRdB}).}\label{fig:capacities_optical}
\end{figure}

The capacity curve \cite[Fig.~36, Curve (3)]{Essiambre2010} was produced using a 16-ring input.
Instead of using (\ref{eq:partially_coherent_channel_discrete_time_spectralloss_capacity}), which holds for a bidimensional Gaussian input, we calculate the polar decomposition's amplitude and phase terms for the channel model (\ref{eq:partially_coherent_channel_discrete_time_spectralloss}) with a 16-ring input.
The (very small) mixed~term~II is neglected.
A good fit of the resulting capacity curve with \cite[Fig.~36, Curve (3)]{Essiambre2010} is obtained for $c \!=\! 1.1 \cdot 10^{5} {\rm ~W}^{-2}$, see Fig.~\ref{fig:capacities_optical}.
The WDM system capacity curve \cite[Fig.~36, Curve (1)]{Essiambre2010} is shown in red for reference.
We observe that the channel model (\ref{eq:partially_coherent_channel_discrete_time_spectralloss}) reproduces the sharp capacity decline in the high-power region well.
However, the spectral loss model (\ref{eq:partially_coherent_channel_discrete_time_spectralloss}) with $\sigma^2 \!=\! c \cdot P_{\rm s}^2$ exhibits a sharp capacity roll-off that does not match the shape of the WDM curve shown in Fig.~\ref{fig:capacities_optical}.
This model of spectral loss is clearly insufficient to explain the WDM curve and additional investigations are needed to find mechanisms that would reproduce the WDM curve. 

Finally, we would like to mention that a reviewer of this paper pointed out a discrepancy between the numerical results \cite{Essiambre2010} and an analytical model with phase noise and spectral loss (such as (\ref{eq:partially_coherent_channel_discrete_time_spectralloss_plus_PN}) or (\ref{eq:partially_coherent_channel_discrete_time_spectralloss})) if the noise $N$ is set to zero.
In this case, the information rate for large but finite $P_{\rm s}$ will clearly be $\log_2(r)$ where $r$ is the number of rings.
The capacity therefore does not reduce with $P_{\rm s}$.
This is not supported by the results in \cite[Fig.~36, Curve (2)]{Essiambre2010} and it shows that spectral loss cannot completely account for the capacity reduction at high signal power.
This is especially apparent if the ASE noise power is small.
Thus, as emphasized above, spectral loss should be considered as only \emph{one} mechanism by which fiber capacity can exhibit a maximum and approach zero at high signal powers.

\section{Conclusion}\label{sec:conclusion}
We have presented a polar decomposition of the mutual information between a complex-valued channel input and its output.
This decomposition yields two main terms, an amplitude term and a phase term, and two ``mixed'' terms that are small or zero in many cases.
The decomposition was performed for the AWGN channel with a Gaussian input (for which asymptotic analytical approximations are derived), a phase-modulated input, and with discrete input constellations.

Partially coherent channels are channels with AWGN and additional phase noise.
The decomposition amplitude term of such channels is not affected by phase noise.
In contrast, the decomposition phase term is bounded because of phase noise.
A property of partially coherent channels with white phase noise that we call spectral loss was derived and discussed.
Effectively, this loss decreases the received SNR; hence, the decomposition amplitude term is affected by phase noise, too.
Spectral loss must be taken into account in the analysis of channels impaired by phase noise as well as in their numerical simulation.
A particularly interesting example of a partially coherent channel is the nonlinear fiber-optic channel.
Capacity results for optical channels limited by signal-noise interaction were calculated.

Finally, the polar decomposition is useful to understand the fundamental impairments of channels such as partially coherent channels and their optimizing input constellations.
The decomposition is a practical tool for a rapid numerical evaluation of mutual information in cases where the mixed terms are small and the complex-valued channel can be effectively decomposed into two independent one-dimensional channels.

\section*{Acknowledgment}
The authors would like to acknowledge inspiring discussions with Gerard J. Foschini on various aspects of the problems treated in this paper, in particular on the nature and treatment of phase noise.

\appendices
\section{Review of directional statistics}\label{app:directional_statistics}
Random variables such as phase angles or points on a spherical surface cannot be treated with ``conventional'' statistical methods. (E.g., the average wind direction calculated from two measurements of $358\degree$ and $2\degree$ is not $180\degree$.)
The field that deals with such \emph{directional} (in contrast to \emph{linear}) random variables is known as \emph{directional statistics}~\cite{Mardia1972}.

\subsection{Trigonometric moments}
We restrict our review to one-dimensional directional (or circular) random variables, \eg, phase angles.
Such a random variable $\Theta$ is defined on an arbitrary interval of length $2\pi$ and has a periodic probability density function (PDF) that satisfies
\begin{equation}
\int_{c-\pi}^{c+\pi} p(\theta) \dd{\theta} = 1, \quad c \in \mathbb{R}.
\label{eq:PDF_directional_rv}
\end{equation}
To ensure that the statistical moments of the directional random variable are invariant under a rotation of the coordinate system, the \emph{trigonometric moments} are calculated from $e^{\jm\Theta}$ rather than from $\Theta$.
The $i^{\rm th}$ trigonometric moment of $\Theta$ is defined as~\cite{Fisher1996}
\begin{equation}
\int_{-\pi}^{\pi} \left( e^{\jm\theta} \right)^i p(\theta) \dd{\theta}.
\label{eq:definition_trigonometric_moment}
\end{equation}
The first trigonometric moment can be calculated as
\begin{equation}
\int_{-\pi}^{\pi} e^{\jm\theta} p(\theta) \dd{\theta} = \rho^\circ_\Theta \cdot e^{\jm\mu^\circ_\Theta},
\label{eq:first_trigonometric_moment}
\end{equation}
where $\rho^\circ_\Theta$ is the \emph{resultant length} and $\mu^\circ_\Theta$ is the \emph{mean direction} of $\Theta$~\cite{Mardia1972}.
The $i^{\rm th}$ \emph{central trigonometric moment} is calculated as the $i^{\rm th}$ trigonometric moment of $\Theta-\mu^\circ_\Theta$.

To quantify the \emph{concentration} (or, inversely, the \emph{dispersion}) of a circular random variable $\Theta$, it is common to define the \emph{circular variance} as~\cite{Mardia1972,Fisher1996}
\begin{equation}
V^\circ_\Theta = 1 - \left| \expectation{}{e^{\jm\Theta}} \right| = 1 - \rho^\circ_\Theta.
\label{eq:circular_variance}
\end{equation}
Clearly, the circular variance is maximized if $\Theta$ is uniformly distributed ($V^\circ_\Theta=1$) and minimized for a constant $\Theta$ ($V^\circ_\Theta=0$).
It must be noted that the \emph{circular standard deviation} is \emph{not} defined as $\sqrt{V^\circ_\Theta}$, but as~\cite{Fisher1996}
\begin{equation}
\sigma^\circ_\Theta = \sqrt{-2\ln(1-V^\circ_\Theta)} = \sqrt{-2\ln \rho^\circ_\Theta}.
\label{eq:circular_std_deviation}
\end{equation}

\subsection{Circular distributions}
An example for a circular distribution has been introduced above in (\ref{eq:phase_PDF_AWGN}), which describes the probability density of the phase angle of a complex phasor corrupted by complex-valued AWGN.
This distribution ranges from a uniform distribution (in any $2\pi$ interval) for small SNRs to a Gaussian distribution for large SNRs.
Middleton gives a series expansion of (\ref{eq:phase_PDF_AWGN})~\cite[\S~9.2-2]{Middleton} which has been applied in the context of systems with phase noise (\cf references given in \cite[Appendix~4.A]{Ho_PM_systems_2005}).

\subsubsection{Wrapped Gaussian distribution}
Another important circular distribution is the \emph{wrapped Gaussian} distribution~\cite{Kotz1982,Bahlmann2006}:
\begin{equation}
p(\theta) = \frac{1}{\sqrt{2\pi}\sigma} \cdot \sum_{k=-\infty}^{\infty}
	\exp\left( -\frac{(\theta-\mu-2\pi k)^2}{2\sigma^2} \right).
\label{eq:wrapped_Gaussian}
\end{equation}
This distribution occurs when a linear random variable $X \sim \normalr{\mu}{\sigma^2}$ is ``wrapped'' around a circle, \ie, $\Theta = X \mod 2\pi$.

The mean direction $\mu^\circ_\Theta$, resultant length $\rho^\circ_\Theta$ and circular variance $V^\circ_\Theta$ of a wrapped Gaussian random variable can be calculated as~\cite{Bahlmann2006}
\begin{equation}
\mu^\circ_\Theta = \mu \!\! \mod 2\pi, \quad \rho^\circ_\Theta = e^{-\frac{1}{2} \sigma^2}
\label{eq:wrapped_Gaussian_resultantlength}
\end{equation}
and
\begin{equation}
V^\circ_\Theta = 1 - e^{-\frac{1}{2} \sigma^2}. 
\label{eq:wrapped_Gaussian_circularvariance}
\end{equation}
The wrapped Gaussian approaches a uniform distribution for large $\sigma$ and can be approximated by a Gaussian distribution for small $\sigma$ as shown in Fig.~\ref{fig:wG_vM_PDFs}.

\subsubsection{Von Mises distribution}
While the wrapped Gaussian distribution shares some of the properties of the linear Gaussian distribution~\cite{Rehacek2008}, it does not maximize the entropy for a given (circular) variance.
This condition is met by the \emph{von Mises} distribution~\cite{Kotz1982,Fisher1996}
\begin{equation}
p(\theta) = \frac{\exp(\kappa\cos(\theta-\mu))}{2 \pi \Besseli_0(\kappa)},
\label{eq:vonMises}
\end{equation}
where $\mu$ is the circular mean (and is usually called the \emph{centrality parameter}), $\kappa$ is the \emph{concentration parameter} and $\Besseli_0(.)$ is the modified Bessel function of the first kind with order zero.
In engineering, the von Mises distribution is known as the \emph{Tikhonov} distribution (after V.~I.~Tikhonov)~\cite{Abreu2007}; it appears in the description of the phase error of phase-locked loops~\cite{Viterbi1963}.

The circular variance is calculated using (\ref{eq:circular_variance}) with (\ref{eq:first_trigonometric_moment}) as
\begin{align}
V^\circ_\Theta &= 1 - \rho^\circ_\Theta \nonumber \\
 &= 1 - \left| \int_{-\pi}^{\pi} p(\theta) \cdot e^{\jm\theta} \dd{\theta} \right| \nonumber \\
 &= 1 - \frac{1}{2 \pi \Besseli_0(\kappa)} \left|
 	\int_{-\pi}^{\pi} e^{\kappa\cos\theta} (\cos\theta + \jm\sin\theta) \dd{\theta} \right| \nonumber \\
 &= 1 - \frac{1}{\pi \Besseli_0(\kappa)} \int_{0}^{\pi} e^{\kappa\cos\theta} \cos\theta \dd{\theta}
 	= 1 - \frac{\Besseli_1(\kappa)}{\Besseli_0(\kappa)}.
\label{eq:vonMises_circular_variance}
\end{align}
To obtain (\ref{eq:vonMises_circular_variance}), we use the modified Bessel functions of the first kind of order $n$ defined as (see~\cite{Abramowitz})
\begin{equation}
\Besseli_n(\kappa) = \frac{1}{\pi} \int_{0}^{\pi} e^{\kappa \cos x} \cos(nx) \dd{x}.
\label{eq:Bessel_function_identity}
\end{equation}

The differential entropy is calculated as
\begin{align}
h(\Theta) &= \int_{-\pi}^{\pi} \frac{e^{\kappa\cos\theta}}{2 \pi \Besseli_0(\kappa)}
		\ln \frac{2 \pi \Besseli_0(\kappa)}{e^{\kappa\cos\theta}} \dd{\theta} \nonumber \\
 &= \frac{\ln(2\pi \Besseli_0(\kappa))}{2\pi \Besseli_0(\kappa)} \cdot \int_{-\pi}^{\pi} e^{\kappa\cos\theta} \dd{\theta}
 		- \frac{k}{2\pi \Besseli_0(\kappa)} \cdot \int_{-\pi}^{\pi} e^{\kappa\cos\theta} \cos\theta \dd{\theta} \nonumber \\
 &= \ln(2\pi \Besseli_0(\kappa)) - k \cdot \frac{\Besseli_1(\kappa)}{\Besseli_0(\kappa)},
\label{eq:vonMises_entropy}
\end{align}
where (\ref{eq:Bessel_function_identity}) was used twice in the last equality.

Among all linear distributions that satisfy an average power (or variance) constraint $\expectation{}{|X|^2} \leq P$, the Gaussian distribution maximizes the differential entropy $h(X)$~\cite{CoverThomas}.
Similarly, one can ask for the circular distribution $p(\theta)$ that maximizes $h(\Theta)$ under a circular variance constraint $V^\circ_\Theta \leq A$.
Without loss of generality, we assume $\mu^\circ_\Theta=0$ which means that $\expectation{}{e^{\jm\Theta}}$ is a non-negative real number and that $\expectation{}{\sin\Theta}=0$.
We can thus write the circular variance constraint as
\begin{align}
V^\circ_\Theta &\stackrel{(\ref{eq:circular_variance})}{=}
		1 - |\expectation{}{e^{\jm\Theta}}| \nonumber \\
 &= 1 - \int_{-\pi}^{\pi} p(\theta) \cos\theta \dd{\theta} - \jm \underbrace{\int_{-\pi}^{\pi} p(\theta) \sin\theta \dd{\theta}}_{=0} \nonumber \\
 &= 1 - \expectation{}{\cos\Theta} \leq A.
\label{eq:circular_variance_constraint}
\end{align}

To prove that the von Mises distribution (\ref{eq:vonMises}) maximizes the differential entropy under the circular variance contraint~(\ref{eq:circular_variance_constraint}), we calculate the Kullback-Leibler distance between the von Mises distribution $p(\theta)$ and an arbitrary other distribution $q(\theta)$:
\begin{align}
D(q \Vert p) &= \int_{-\pi}^{\pi} q(\theta) \ln \frac{q(\theta)}{p(\theta)} \dd{\theta} \nonumber \\
 &= \underbrace{\int_{-\pi}^{\pi} q(\theta) \ln q(\theta) \dd{\theta}}_{-h(q)}
			- \int_{-\pi}^{\pi} q(\theta) \ln p(\theta) \dd{\theta} \nonumber \\
 &= -h(q) - \int_{-\pi}^{\pi} q(\theta)
 			\ln \frac{e^{\kappa\cos\theta}}{2 \pi \Besseli_0(\kappa)} \dd{\theta} \nonumber \\
 &= -h(q) + \ln(2\pi \Besseli_0(\kappa)) - \kappa \cdot \int_{-\pi}^{\pi} q(\theta)
 			\cos\theta \dd{\theta} \nonumber \\
 &= -h(q) + \ln(2\pi \Besseli_0(\kappa)) - \kappa \cdot \underbrace{\expectation{\Theta\sim q(\theta)}{\cos\Theta}}_{\geq 1-A} \nonumber \\
 &\leq -h(q) + h(p),
\label{eq:vonMises_max_entropy_proof}
\end{align}
where $h(q)$ denotes the differential entropy $h(\Theta)$ of a random variable $\Theta \sim q(\theta)$ and where $\kappa$ is chosen to satisfy $1-A=\Besseli_1(\kappa)/\Besseli_0(\kappa)$.
Recall that $D(q \Vert p) \geq 0$ with equality if and only if $p \!=\! q$ \cite{CoverThomas}.
Hence, we find that
\begin{equation}
h(p) \geq h(q)
\label{eq:vonMises_max_entropy_proof_3}
\end{equation}
with equality if and only if $q=p$.

A different path to get to the same result is to note that the von Mises distribution is a special case of the maximum entropy distribution \cite[p.~267]{CoverThomas}.
With the constraint (\ref{eq:circular_variance_constraint}), the maximum entropy distribution with coefficients $\lambda_0 \!=\! -\ln(2\pi \Besseli_0(\kappa))$ and $\lambda_1=\kappa$ transforms into (\ref{eq:vonMises}).
Barakat finds the same result using Lagrange multipliers~\cite{Barakat}.
Observe that the von Mises distribution becomes uniform for large circular variance (small $\kappa$) and approaches a Gaussian distribution with variance $\sigma^2=1/\kappa$ when the circular variance is small ($\kappa$ large)~\cite{Kotz1982}.
Fig.~\ref{fig:wG_vM_PDFs} shows the wrapped Gaussian PDF for $\mu=0$ (\ie, $\mu^\circ_\Theta=0$) and various values of $\sigma$.

Because of its maximum entropy property, the von Mises distribution is often considered to be the circular analogue of the linear normal distribution.
Hence, it is sometimes referred to as the \emph{circular normal distribution}; to avoid confusion with the wrapped Gaussian distribution, it is advisable not to use this term.
The wrapped Gaussian and the von Mises distribution have a very similar shape~\cite{Barakat}, see Fig.~\ref{fig:wG_vM_PDFs}.
In practice, one often uses whichever is more convenient~\cite{Fisher1996}.

\subsubsection{Truncated Gaussian distribution}
Suppose now for the sake of argument that the phase constraint is the usual second-order constraint $\expectation{}{\Theta^2} \leq A$, where the expectation is performed over the interval $[-\pi,\pi)$.
Suppose further that we wish to maximize the entropy PDF over all PDFs with $\expectation{}{\Theta}=0$ (the latter constraint is made to simplify the discussion).
Consider the \emph{truncated} Gaussian distribution
\begin{equation}
p(\theta) = \frac{\lambda}{\sqrt{2\pi}\sigma}
		\exp\left( - \frac{\theta^2}{2 \sigma^2} \right), \quad -\pi \leq \theta < \pi,
\end{equation}
where $\lambda$ is a scaling constant that ensures (\ref{eq:PDF_directional_rv}) is valid, and $\sigma^2$ is chosen so that $\expectation{}{\Theta^2}=A$.
We compute
\begin{align}
h(\Theta) &= \int_{-\pi}^{\pi} - p(\theta) \ln p(\theta) \dd{\theta} \nonumber \\
 &= \frac{1}{2} \ln\left( \frac{2 \pi \sigma^2}{\lambda^2} \right) + \frac{1}{2 \sigma^2} \cdot
		\underbrace{\int_{-\pi}^{\pi}  p(\theta) \theta^2 \dd{\theta}}_{=\expectation{}{\Theta^2}=A} \nonumber \\
 &= \frac{1}{2} \ln\left( \frac{2 \pi \sigma^2}{\lambda^2} \right) + \frac{A}{2 \sigma^2}.
\end{align}
We further have
\begin{align}
D(q \Vert p) &= \int_{-\pi}^{\pi} q(\theta) \ln \frac{q(\theta)}{p(\theta)} \dd{\theta} \nonumber \\
 &= -h(q) - \int_{-\pi}^{\pi} q(\theta) \ln p(\theta) \dd{\theta} \nonumber \\
 &= -h(q) + \frac{1}{2} \ln\left( \frac{2\pi\sigma^2}{\lambda^2} \right) +
 			\frac{1}{2\sigma^2} \cdot \int_{-\pi}^{\pi} q(\theta) \theta^2 \dd{\theta} \nonumber \\
 &= -h(q) + \frac{1}{2} \ln\left( \frac{2\pi\sigma^2}{\lambda^2} \right) +
 			\frac{1}{2\sigma^2} \cdot \underbrace{\expectation{\Theta \sim q(\theta)}{\Theta^2}}_{\leq A} \nonumber \\
 &\leq -h(q) + h(p).
\end{align}
Using $D(q \Vert p) \geq 0$ with equality if and only if $q=p$, we find that a truncated Gaussian distribution maximizes entropy.

\begin{figure}[!htb]
\centering
  \psfrag{x}[][]{$\theta/\pi$}
  \psfrag{y}[][]{$p(\theta)$}
  \psfrag{s05}{$\sigma=0.5$}		
  \psfrag{s1}{$\sigma=1$}
  \psfrag{s2}{$\sigma=2$}
  \psfrag{s5}{$\sigma=5$}
  \includegraphics[width=0.7\textwidth]{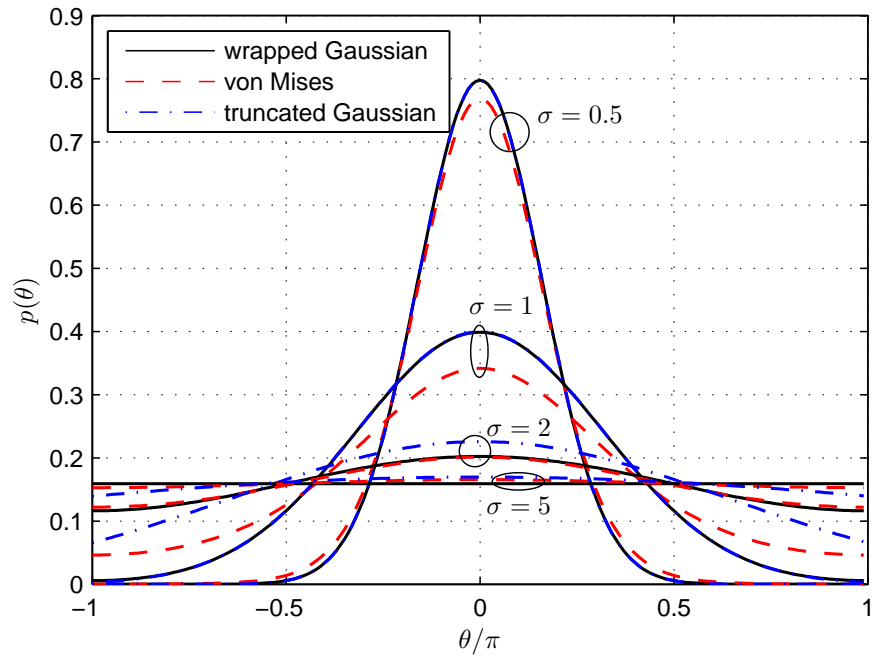}
\caption{Wrapped and truncated Gaussian and von Mises PDFs with $\mu=0$ and various values of $\sigma$ and $\kappa=1/\sigma^2$.}\label{fig:wG_vM_PDFs}
\end{figure}

Fig.~\ref{fig:wG_vM_PDFs} shows the PDFs for the truncated Gaussian distribution for $\expectation{}{\Theta}=0$ and various values of $\sigma$ (wrapped and truncated Gaussians) and $\kappa=1/\sigma^2$ (von Mises).
We remark that the physical meaning of our second-order constraint is unclear, but the same can be said for the circular variance constraint.
It is interesting, however, that maximum entropy considerations lead to either a von Mises distribution or a truncated Gaussian distribution.
Two interesting problems are whether the wrapped Gaussian distribution is maximum-entropy under some natural circular constraint, and whether the wrapped Gaussian has other natural ``normal'' properties~\cite{Rehacek2008}.


\bibliographystyle{IEEEtran}
\bibliography{IEEEabrv,bernhards_literatur}

\begin{thebibliography}{10}
\providecommand{\url}[1]{#1}
\csname url@samestyle\endcsname
\providecommand{\newblock}{\relax}
\providecommand{\bibinfo}[2]{#2}
\providecommand{\BIBentrySTDinterwordspacing}{\spaceskip=0pt\relax}
\providecommand{\BIBentryALTinterwordstretchfactor}{4}
\providecommand{\BIBentryALTinterwordspacing}{\spaceskip=\fontdimen2\font plus
\BIBentryALTinterwordstretchfactor\fontdimen3\font minus
  \fontdimen4\font\relax}
\providecommand{\BIBforeignlanguage}[2]{{%
\expandafter\ifx\csname l@#1\endcsname\relax
\typeout{** WARNING: IEEEtran.bst: No hyphenation pattern has been}%
\typeout{** loaded for the language `#1'. Using the pattern for}%
\typeout{** the default language instead.}%
\else
\language=\csname l@#1\endcsname
\fi
#2}}
\providecommand{\BIBdecl}{\relax}
\BIBdecl

\bibitem{Cahn1960}
C.~Cahn, ``Combined digital phase and amplitude modulation communication
  systems,'' \emph{IRE Transactions on Communications Systems}, vol.~8, no.~3,
  pp. 150--155, Sep. 1960.

\bibitem{Hanzo_QAM}
L.~Hanzo, W.~Webb, and T.~Keller, \emph{Single-- and multi-carrier quadrature
  amplitude modulation}.\hskip 1em plus 0.5em minus 0.4em\relax Chichester:
  John Wiley \& Sons, Inc., 2000.

\bibitem{Blachman1953}
N.~M. Blachman, ``A comparison of the informational capacities of amplitude-
  and phase-modulation communication systems,'' \emph{Proceedings of the IRE},
  vol.~41, no.~6, pp. 748--759, Jun. 1953.

\bibitem{Viterbi1963}
A.~J. Viterbi, ``Phase-locked loop dynamics in the presence of noise by
  {F}okker-{P}lanck techniques,'' \emph{Proc. {IEEE}}, vol.~51, no.~12, pp.
  1737--1753, Dec. 1963.

\bibitem{Foschini1973}
G.~J. Foschini, R.~D. Gitlin, and S.~B. Weinstein, ``On the selection of a
  two-dimensional signal constellation in the presence of phase jitter and
  {G}aussian noise,'' \emph{Bell System Technical Journal}, vol.~52, no.~6, pp.
  927--965, Jul.-Aug. 1973.

\bibitem{Katz2004}
M.~Katz and S.~Shamai~(Shitz), ``On the capacity-achieving distribution of the
  discrete-time noncoherent and partially coherent {AWGN} channels,''
  \emph{{IEEE} Trans. Inf. Theory}, vol.~50, no.~10, pp. 2257--2270, Oct. 2004.

\bibitem{CoverThomas}
T.~Cover and J.~Thomas, \emph{Elements of Information Theory}.\hskip 1em plus
  0.5em minus 0.4em\relax New York: John Wiley \& Sons, Inc., 1991.

\bibitem{Proakis}
J.~G. Proakis, \emph{Digital communications}, 3rd~ed.\hskip 1em plus 0.5em
  minus 0.4em\relax New York: McGraw-Hill, 1995.

\bibitem{Abramowitz}
M.~Abramowitz and I.~A. Stegun, Eds., \emph{Handbook of mathematical
  functions}, 10th~ed.\hskip 1em plus 0.5em minus 0.4em\relax Washington:
  United States Government Printing Office, 1972.

\bibitem{Middleton}
D.~Middleton, \emph{An introduction to statistical communication theory}.\hskip
  1em plus 0.5em minus 0.4em\relax New York: McGraw-Hill, 1960.

\bibitem{Aldis1993}
J.~P. Aldis and A.~G. Burr, ``The channel capacity of discrete time phase
  modulation in {AWGN},'' \emph{{IEEE} Trans. Inf. Theory}, vol.~39, no.~1, pp.
  184--185, Jan. 1993.

\bibitem{Blahut1987}
R.~E. Blahut, \emph{Principles and practice of information theory}.\hskip 1em
  plus 0.5em minus 0.4em\relax Reading, MA: Addison-Wesley, 1987.

\bibitem{Geist1990}
J.~M. Geist, ``Capacity and cutoff rate for dense {M}-ary {PSK}
  constellations,'' in \emph{Proceedings of the Military Communications
  Conference (MILCOM)}, vol.~2, Sep. 30 -- Oct. 3, 1990, pp. 768--770.

\bibitem{HoKahn2002}
K.-P. Ho and J.~M. Kahn, ``Channel capacity of {WDM} systems using
  constant-intensity modulation formats,'' in \emph{Proceedings of the Optical
  Fiber Communication Conference (OFC)}, Mar. 17--22, 2002, pp. 731--733.

\bibitem{Belzer2002}
B.~J. Belzer, A.~D. Risley, P.~Hou, and T.~R. Fischer, ``Capacity of {AM-PSK}
  on partially coherent fading channels,'' \emph{{IEEE} Trans. Commun.},
  vol.~50, no.~3, pp. 369--373, Mar. 2002.

\bibitem{Agrawal}
G.~P. Agrawal, \emph{Fiber-optic communication systems}, 3rd~ed.\hskip 1em plus
  0.5em minus 0.4em\relax New York: John Wiley \& Sons, Inc., 2002.

\bibitem{Essiambre2010}
R.-J. Essiambre, G.~Kramer, P.~J. Winzer, G.~J. Foschini, and B.~Goebel,
  ``Capacity limits of optical fiber networks,'' \emph{J. Lightw. Technol.},
  vol.~28, no.~4, pp. 662--701, Feb. 2010.

\bibitem{Viterbi1965}
A.~Viterbi, ``Optimum detection and signal selection for partially coherent
  binary communication,'' \emph{{IEEE} Trans. Inf. Theory}, vol.~11, no.~2, pp.
  239--246, Apr. 1965.

\bibitem{Born_Wolf}
M.~Born and E.~Wolf, \emph{Principles of optics}, 7th~ed.\hskip 1em plus 0.5em
  minus 0.4em\relax Cambridge: Cambridge University Press, 1999.

\bibitem{Salz1985}
J.~Salz, ``Coherent lightwave communication,'' \emph{AT\&T Technical Journal},
  vol.~64, no.~10, pp. 2153--2209, Dec. 1985.

\bibitem{Foschini1988}
G.~J. Foschini, L.~J. Greenstein, and G.~Vannucci, ``Noncoherent detection of
  coherent lightwave signals corrupted by phase noise,'' \emph{{IEEE} Trans.
  Commun.}, vol.~36, no.~3, pp. 306--314, Mar. 1988.

\bibitem{Foschini1989}
G.~J. Foschini, G.~Vannucci, and L.~J. Greenstein, ``Envelope statistics for
  filtered optical signals corrupted by phase noise,'' \emph{{IEEE} Trans.
  Commun.}, vol.~37, no.~12, pp. 1293--1302, Dec. 1989.

\bibitem{Shamai1989}
S.~Shamai~(Shitz), ``On information rates of bandwidth restricted noisy phase
  channels,'' \emph{Archiv f{\"u}r Elektronik und {\"U}bertragungstechnik
  (Electronics and Communication)}, vol.~43, no.~6, pp. 350--360, Nov.--Dec.
  1989.

\bibitem{Dallal1991}
Y.~E. Dallal and S.~Shamai~(Shitz), ``An upper bound on the error probability
  of quadratic-detection in noisy phase channels,'' \emph{{IEEE} Trans.
  Commun.}, vol.~39, no.~11, pp. 1635--1650, Nov. 1991.

\bibitem{AgrawalNL}
G.~P. Agrawal, \emph{Nonlinear fiber optics}, 4th~ed.\hskip 1em plus 0.5em
  minus 0.4em\relax Burlington, MA: Academic Press, 2007.

\bibitem{Ho_PM_systems_2005}
K.-P. Ho, \emph{Phase-modulated optical communication systems}.\hskip 1em plus
  0.5em minus 0.4em\relax New York: Springer, 2005.

\bibitem{Peleg2000}
M.~Peleg, S.~Shamai~(Shitz), and S.~Gal{\'a}n, ``Iterative decoding for coded
  noncoherent {MPSK} communications over phase-noisy {AWGN} channel,''
  \emph{IEE Proceedings-Communications}, vol. 147, no.~2, pp. 87--95, Apr.
  2000.

\bibitem{Gray1995}
R.~M. Gray and J.~W. Goodman, \emph{Fourier transforms}.\hskip 1em plus 0.5em
  minus 0.4em\relax Boston: Kluwer Academic Publishers, 1995.

\bibitem{Jeruchim2000}
M.~C. Jeruchim, P.~Balaban, and K.~S. Shanmugan, \emph{Simulation of
  communication systems}, 2nd~ed.\hskip 1em plus 0.5em minus 0.4em\relax New
  York: Kluwer Academic / Plenum Publishers, 2000.

\bibitem{Hou2002}
P.~Hou, B.~J. Belzer, and T.~R. Fischer, ``Shaping gain of the partially
  coherent additive white {G}aussian noise channel,'' \emph{{IEEE} Commun.
  Lett.}, vol.~6, no.~5, pp. 175--177, May 2002.

\bibitem{Hou2003}
------, ``On the capacity of the partially coherent additive white {G}aussian
  noise channel,'' in \emph{Proceedings of the IEEE International Symposium on
  Information Theory}, Jun. 29--Jul. 4, 2003, p. 372.

\bibitem{Hranilovic2004}
S.~Hranilovic and F.~R. Kschischang, ``Capacity bounds for power- and
  band-limited optical intensity channels corrupted by {G}aussian noise,''
  \emph{{IEEE} Trans. Inf. Theory}, vol.~50, no.~5, pp. 784--795, May 2004.

\bibitem{Colavolpe2001}
G.~Colavolpe and R.~Raheli, ``The capacity of the noncoherent channel,''
  \emph{European Transactions on Telecommunications}, vol.~12, no.~4, pp.
  289--296, Jul.-Aug. 2001.

\bibitem{Ho2005}
K.-P. Ho, ``Exact evaluation of the capacity for intensity-modulated
  direct-detection channels with optical amplifier noises,'' \emph{{IEEE}
  Photon. Technol. Lett.}, vol.~17, no.~4, pp. 858--860, Apr. 2005.

\bibitem{Mecozzi2001}
A.~Mecozzi and M.~Shtaif, ``On the capacity of intensity modulated systems
  using optical amplifiers,'' \emph{{IEEE} Photon. Technol. Lett.}, vol.~13,
  no.~9, pp. 1029--1031, Sep. 2001.

\bibitem{Mao2008}
W.~Mao and J.~M. Kahn, ``Lattice codes for amplified direct-detection optical
  systems,'' \emph{{IEEE} Trans. Commun.}, vol.~56, no.~7, pp. 1137--1145, Jul.
  2008.

\bibitem{OhmLueke}
J.-R. Ohm and H.~D. L{\"u}ke, \emph{Signal{\"u}bertragung}, 8th~ed.\hskip 1em
  plus 0.5em minus 0.4em\relax Berlin: Springer, 2002.

\bibitem{Mardia1972}
K.~V. Mardia, \emph{Statistics of Directional Data}.\hskip 1em plus 0.5em minus
  0.4em\relax New York: Academic Press, 1972.

\bibitem{Fisher1996}
N.~I. Fisher, \emph{Statistical Analysis of Circular Data}.\hskip 1em plus
  0.5em minus 0.4em\relax Cambridge: Cambridge University Press, 1996.

\bibitem{Kotz1982}
S.~Kotz and N.~L. Johnson, Eds., \emph{Encyclopedia of statistical
  sciences}.\hskip 1em plus 0.5em minus 0.4em\relax New York: John Wiley \&
  Sons, Inc., 1982, vol.~2.

\bibitem{Bahlmann2006}
C.~Bahlmann, ``Directional features in online handwriting recognition,''
  \emph{Pattern Recognition}, vol.~39, no.~1, pp. 115--125, Jan. 2006.

\bibitem{Rehacek2008}
J.~\v{R}eh\'a\v{c}ek, Z.~Bouchal, R.~\v{C}elechovsk\'y, Z.~Hradil, and L.~L.
  S\'anchez-Soto, ``Experimental test of uncertainty relations for quantum
  mechanics on a circle,'' \emph{Physical Review A}, vol.~77, no.~3, p. 032110,
  Mar. 2008.

\bibitem{Abreu2007}
G.~T.~F. de~Abreu, ``On the simulation of {T}ikhonov random processes,'' in
  \emph{Proceedings of the IEEE International Conference on Communications
  (ICC)}, Jun. 24--28, 2007, pp. 5021--5027.

\bibitem{Barakat}
R.~Barakat, ``Weak-scatterer generalization of the {K}-density function. {II}.
  {P}robability density of total phase,'' \emph{Journal of the Optical Society
  of America A}, vol.~4, no.~7, pp. 1213--1219, Jul. 1987.

\end{thebibliography}

\begin{IEEEbiographynophoto}{Bernhard Goebel}
(S'04-M'10) received the Dipl.-Ing. degree in electrical engineering and information technology in 2004 and the Dr.-Ing. degree in 2010, both from Technische Universit{\"a}t M{\"u}nchen (TUM), Munich, Germany. His diploma thesis was on the information-theoretic analysis of complex genetic diseases. The topic of his doctoral dissertation were information-theoretic aspects of nonlinear fiber-optic communication systems. Dr. Goebel spent graduate semesters at the University of Southampton, U.K., and Siemens Corporate Research, Princeton, NJ, where he worked on segmentation algorithms for medical imaging. From 2004 to 2010, he was a member of the technical staff at TUM's Institute for Communications Engineering. On leave from TUM, he spent the summer of 2009 as a visiting researcher at Alcatel-Lucent Bell Laboratories, Holmdel, NJ. In 2010, he served as head of the organization committee of the International Conference on Transparent Optical Networks (ICTON) held in Munich. Since 2011, he has been with the concept development division of Volkswagen, Wolfsburg, Germany. Dr. Goebel has published in the fields of medical imaging, bioinformatics, optical communications and information theory, and is the inventor of two patents. He is a member of the IEEE and VDE/ITG.
\end{IEEEbiographynophoto}

\begin{IEEEbiographynophoto}{Ren\'e-Jean Essiambre}
(S'95-SM'06) received the B.S. and Ph.D. degrees in physics and optics from Université Laval, Quebec City, QC, Canada. During his Ph.D. studies, he spent one year at McGill University, Montreal, QC, Canada, where he was engaged in research on solid-state physics. From 1995 to 1997, he was a Postdoctoral Fellow with Prof. Agrawal at The Institute of Optics, University of Rochester, Rochester, NY. Since 1997, he has been at Bell Laboratories, Alcatel-Lucent, Holmdel, NJ. His early research has been focused on optical switching, soliton communication systems, high-power fiber lasers, and mode-locked fiber lasers. His current research interests include high-speed transmission (100 Gb/s and above) and physical layer design of fiber-optic communication systems, including Raman amplification, Rayleigh backscattering, fiber nonlinearities, network design, advanced modulation formats, information theory, and coding. He is the author and coauthor of more than 100 scientific publications and several book chapters. He has served on many conference committees including ECOC, OFC, CLEO, and LEOS. He is program co-chair of CLEO: Science and Innovations 2012. Dr. Essiambre is a Fellow of the Optical Society of America (OSA). He is the recipient of the 2005 OSA Engineering Excellence Award. He is also a Distinguished Member of Technical Staff at Bell Laboratories.
\end{IEEEbiographynophoto}

\begin{IEEEbiographynophoto}{Gerhard Kramer}
(S'91-M'94-SM'08-F'10) received the B.Sc. and M.Sc. degrees in electrical engineering from the University of Manitoba, Winnipeg, MB, Canada in 1991 and 1992, respectively, and the Dr. sc. techn. degree from the Swiss Federal Institute of Technology (ETH), Z\"urich, Switzerland, in 1998.

From 1998 to 2000, he was with Endora Tech AG, Basel, Switzerland. From 2000 to 2008, he was with Bell Laboratories, Alcatel-Lucent, Murray Hill, NJ. From 2009 to 2010 he was a faculty member of the University of Southern California (USC), Los Angeles. Since 2010 he is Alexander von Humboldt Professor at the Technische Universit\"at M\"unchen, Germany.

Prof. Kramer is a member of the Board of Governors of the IEEE Information Theory Society since 2009. He is the Society's 2nd Vice President in 2011. He has served as an Associate Editor, Guest Editor, and Publications Editor for the IEEE TRANSACTIONS ON INFORMATION THEORY. He served as Co-Chair of the Technical Program Committee of the 2008 IEEE International Symposium on Information Theory, and as Founding Co-Chair of the first,
second, and third Annual Schools of Information Theory during 2008-2010. He has been serving as a member of the Emerging Technologies Committee of the IEEE Communications Society since 2009. He is a corecipient of the
IEEE Communications Society 2005 Stephen O. Rice Prize paper award, a Bell Labs President's Gold Award in 2003, and a recipient of an ETH Medal in 1998. He was awarded an Alexander von Humboldt Professorship endowed by the German Federal Ministry of Education and Research in 2010.
\end{IEEEbiographynophoto}

\begin{IEEEbiographynophoto}{Peter J. Winzer}
(S'93-A'99-SM'05-F'09) received the Ph.D. degree in electrical/communications engineering from the Vienna University of Technology, Vienna, Austria, in 1998. His academic work, largely supported by the European Space Agency (ESA), was related to the analysis and modeling of space-borne Doppler wind lidar and highly sensitive free-space optical communication systems. In this context, he specialized on advanced digital optical modulation formats and high sensitivity optical receivers using coherent and direct detection. After joining Bell Labs in November 2000, he focused on various aspects of high-bandwidth optical communication networks, including Raman amplification, optical modulation formats, advanced optical receiver concepts, and digital signal processing at bit rates from 10 to 100-Gb/s. He has widely published in peer-reviewed journals and at conferences and holds several patents in the fields of optical communications, lidar, and data networking. Dr. Winzer is actively involved as a Reviewer, Associate Editor, and Committee Member of various journals and conferences and serves as an elected member of the IEEE-LEOS BoG. He is a Distinguished Member of technical Staff at Bell Labs, a Member of the Optical Society of America (OSA).
\end{IEEEbiographynophoto}

\begin{IEEEbiographynophoto}{Norbert Hanik}
(M'00) was born in 1962. He received the Dipl.-Ing. degree in electrical engineering (with a thesis on digital spread spectrum systems) and the Dr.-Ing. degree (with a dissertation on nonlinear effects in optical signal transmission) from Technische Universit{\"a}t M{\"u}nchen, Munich, Germany, in 1989 and 1995, respectively. He was a Research Associate at TUM's Institute for Telecommunications from 1989 to 1995, where he conducted research in the areas of mobile radio and optical communications. From 1995 to 2004, he was with the Technologiezentrum of Deutsche Telekom, heading the research group System Concepts of Photonic Networks. During his work there, he contributed to a multitude of Telekom internal research and development projects, both as scientist and project leader. During 2002, he was a Visiting Professor at Research Center COM, Technical University of Denmark, Copenhagen. In 2004, he was appointed Associate Professor for wireline and optical transmission systems at TUM. He participated in a large number of research projects, funded by industry, German research funds as well as the European union. Prof. Hanik served as the General Chairman of the International Conference on Transparent Optical Networks (ICTON) 2010 held in Munich. His primary research interests are in the fields of physical design, optimization, operation, and management of optical backbone and access networks. He is a member of IEEE and VDE/ITG.
\end{IEEEbiographynophoto}






\end{document}